\documentclass[12pt,a4paper]{article}
\pdfoutput=1

\usepackage{apalike}
\usepackage{amsmath, amssymb, amsthm}
\usepackage{appendix}
\usepackage{graphicx}

\usepackage[colorlinks = true,linkcolor = blue,urlcolor=blue,hypertexnames=false]{hyperref}
\usepackage[T1]{fontenc}
\usepackage{lmodern} 
\usepackage{newtxtext} 
\usepackage{newtxmath}

\usepackage[shortlabels]{enumitem}
\usepackage[]{setspace} 
\allowdisplaybreaks[1]

\usepackage{csvsimple,longtable,booktabs}

\usepackage[top=4cm, left=3cm, right=3cm]{geometry}
\renewcommand{\theequation}{\arabic{section}.\arabic{equation}}
\counterwithin*{equation}{section}

\begin{document}

\title{The Formation of Global Free Trade Agreement%
\thanks{We would like to thank Jota Ishikawa, Toshiji Miyakawa, Hiroshi Mukunoki, and seminar participants in Game Theory Workshop 2021 and Konan University.  
Shirata acknowledges financial support from the Japan Society for
the Promotion of Science, JSPS KAKENHI Grant Number 18K12738.}
}

\author{
Akira Okada%
\thanks{Professor Emeritus, Hitotsubashi University, 2-1 Naka, Kunitachi, Tokyo 186-8601, JAPAN.
E-mail address: \texttt{aokada313@gmail.com.}} 
and 
Yasuhiro Shirata%
\thanks{Corresponding author. Associate Professor, Department of Economics, Otaru University of Commerce, 3-5-21 Midori, Otaru, Hokkaido, 047-8501, JAPAN. 
E-mail address: \texttt{yasuhiro.shirata@gmail.com}}
}

\date{March, 2021}

\maketitle

\begin{abstract}

We investigate the formation of Free Trade Agreement (FTA) in a competing importers framework with $n$ countries. 
We show that (i) FTA formation causes a negative externality to non-participants, (ii) a non-participant is willing to join an FTA, and (iii) new participation may decrease the welfare of incumbent participants. 
A unique subgame perfect equilibrium of a sequential FTA formation game does not achieve global free trade under an open-access rule where a new applicant needs consent of members for accession, currently employed by many open regionalism agreements including APEC.
We further show that global FTA is a unique subgame perfect equilibrium under an open-access rule without consent.

\vspace{1ex}
\vspace{1ex}


\end{abstract}

\newpage

\section{Introduction}

The rapid proliferation of preferential trade agreements (PTAs) over the last five decades has spurred a large volume of studies on the role of PTAs in the achievement of global free trade. 
The halt of the World Trade Organization (WTO) trade liberalization process in the Doha round has stimulated the growing literature on PTAs.

The literature has extensively explored whether or not PTAs can promote free trade in the world both theoretically and empirically. 
In a recent review, Bagwell et al.\ (2016) write: ``Both the theory and evidence are mixed; hence, as a general matter, whether PTAs are stumbling blocks or building blocks for multilateral liberalization remains ambiguous.'' 
This paper aims to contribute to a debate on the issue by analyzing a formation process to global free trade.

Trade liberalization is a sequential process. 
A small number of countries launch a trade agreement, and the agreement expands with new participants. 
The General Agreement on Tariffs and Trade (GATT) commenced in 1947 with twenty-three countries, and the WTO replace it in 1995. 
Since 2016, the WTO has 164 members. China participated in the WTO in 2001. 
The European Coal and Steel Community commenced in 1950 with six countries, and it has been enlarged through several steps to European Union (EU) with now twenty-seven members. 
The Asia Pacific Economic Cooperation (APEC) started in 1989 with twelve countries, and China, Hong Kong, and Taiwan joined it in 1991. 
As of 2021, it has twenty-one member countries.%
\footnote{WTO: \url{https://www.wto.org/english/thewto_e/history_e/history_e.htm}, EU: \url{https://europa.eu/european-union/about-eu/history_en}, APEC: \url{https://www.apec.org/About-Us/About-APEC/History}, March 14, 2021.}

Since the creation of the North American Free Trade Agreement (NAFTA) in 1994, many PTAs have emerged as alternative vehicles of trade liberalization.\footnote{As of 2021, the cumulative number of RTAs in force has expanded from roughly 50 in 1990 to nearly 350 today (\url{https://rtais.wto.org/UI/charts.aspx}), March 14, 2021.} 
Member countries of PTAs establish jointly internal tariffs on each other. 
The Article XXIV of GATT allows WTO members to form PTAs if the PTAs eliminate tariffs on  ``substantially all trade'' among the member countries and that the external tariffs on non-member countries do not increase as a result of PTA formation. 
While the proponents of PTAs argue that they promote free trade, the proliferation of different PTAs may tangle the world trading system, commonly referred to as a ``spaghetti bowl.'' 
Various commitments and rules of PTAs are overlapped and inconsistent.

Stimulated by the creation of APEC, \emph{open regionalism} has received much attention from academics and policymakers to establish compatibility between regional trading agreements to achieve global free trade. 
The proponents of open regionalism view it as ``a device through which regionalism can be employed to accelerate the progress toward global liberalization and rule-making'' (Bergsten 1997). 
The first definition of open regionalism proposed by Bergsten (1997) is open membership, which requires that ``any country that indicates a credible willingness to accept the rules of the institution would be invited to join.''\footnote{Bergsten (1997) proposes the five definitions. The other four are unconditional MFN, conditional MFN, global liberalization, and trade facilitation.}

The principle of open regionalism, by contrast, has not been appropriately practiced.\footnote{A critic writes that ``APEC's trade liberalisation strategy is a frail initiative'' (Kelegama 2000).} Since Peru, Russia, and Viet Nam joined in 1998, APEC has accepted no new participants. 
Membership is negotiated between applicants and incumbent members.\footnote{The official statement of APEC writes: ``Decisions on the admission of additional members to APEC require a consensus of all existing members'' (21 November 1997, Vancouver).} 
We may regard that the open membership of APEC demands ``consent'' in the sense that any new participant needs unanimous consent of incumbent members. 
To practice open regionalism, Bergsten (1997) and Lewis (2011) argue that APEC should expand the liberalization to all members of the WTO to include more members.\footnote{Bergsten (1997) writes that ``the best, perhaps only, way for APEC to do so is to indicate publicly both its precise liberalisation programme and its willingness to extend that liberalisation to all members of the WTO on a reciprocal basis.''} 

This paper provides a game-theoretic foundation for the recent debate above on open regionalism and free trade. As a framework, we consider a sequential formation process of a free trade agreement (FTA) under an open-access rule with two different participation rules. 
Our international trade model is based on the ``competing importers'' framework of Missios et al.\ (2016).%
\footnote{Missios et al.\ (2016) and Saggi et al.\ (2018) argue that the framework highlights a key insight, \emph{external trade diversion}, that has generally been overlooked in the literature.}

Under the open-access rule with consent that describes the current form of open regionalism of many FTAs, including APEC, a non-member country can participate in an FTA with members' unanimous approval. 
Every country decides sequentially to participate in an FTA according to a fixed order. Whenever a non-member decides to participate in an FTA, all member countries accept or reject new participation independently. An applicant country participates in the FTA if and only if all the incumbents accept it. 
Under the open-access rule without consent, which describes the principle of open regionalism, a non-member can participate in an FTA without an incumbent member's approval.

We summarize the results as follows. We first characterize the equilibrium tariffs under an FTA regime where several overlapping FTAs exist. 
The formation of an FTA does not change the external tariff on non-participants. 
This tariff policy of an FTA is consistent with the GATT Article XXIV, which requires the FTA member countries not to raise their external tariffs on non-members. 
In other words, the GATT Article XXIV is not binding for the external tariff of an FTA.%
\footnote{The optimal internal tariff of an FTA may not be the zero-tariff (Saggi et al. 2019). In this sense, the GATT Article XXIV, which requires the FTA internal tariffs to be zero, is binding in the model.}
Non-participants reduce their tariffs on participants (tariff complementarity, Missios et al. 2016). 
We find that the formation of an FTA affects welfare as follows: 
(i) an FTA causes a negative externality to non-participants, (ii) a non-participant is always willing to join an FTA, and (iii)  new participation may decrease incumbent participants' welfare. 

By these properties, we show that a unique equilibrium FTA under the open-access rule with consent is not the global FTA. 
Incumbent participants reject a new applicant if it exceeds FTA's optimal size, which is strictly smaller than global FTA. 
This result explains that current open regionalism including APEC has not led to global FTA yet, despite of the proponets' expectation. 
By contrast, we show that global FTA is a unique subgame perfect equilibrium under the open-access rule without consent. 

This paper's primary contribution is to show that a participation rule in open regionalism is crucial to achieving the global free trade agreement. 
The strategic behavior of incumbent participants to reject new participants may deteriorate the formation of global free trade. Our result provides theoretical support for the WTO's policy recommendation to require any PTA to be open-access without incumbents' consent.%
\footnote{Several FTAs have open-access provisions. For example, the NAFTA writes: ``any country or group of countries may accede to this Agreement subject to such terms and conditions as may be agreed between such country or countries and the Commission and following approval in accordance with the applicable legal procedures of each country'' (Article 2204). Trans-Pacific Partnership (TPP) Agreement also has an open access provision (Article 30.4).}

The paper is closely related to the existing works of Yi (1996), Seidmann (2009), Mukunoki and Tachi (2006), Missios et al.\ (2016), Furusawa and Konishi (2007), and Goyal and Joshi (2006), all of whom study trade liberalization by game-theoretic models. 
We differentiate our results from these works in the following aspects.%
\footnote{Among other papers on the endogenous PTA formation are Burbidge et al. (1987), Agihon et al. (2007),  Saggi et al. (2010, 2018). The literature employs diverse approaches. In his survey, Maggi (2014) writes: ``this literature can be hard to tame, as the modeling approach and the exact nature of the question seem to shift from paper to paper.''} 

Yi (1996) considers the endogenous formation of custom union (CU) and shows that global free trade is a unique Nash equilibrium in a simultaneous-move open regionalism game.
A critical difference between Yi (1996) and this paper is that there are multiple subgame perfect equilibria in his sequential-move open regionalism game. Global free trade is one of subgame perfect equilibria.%
\footnote{Loke and Winters (2012) present a numerical example which shows that the global free trade result of Yi (1996) under the open regionalism fails under CES preferences and increasing marginal costs in the case of Bertrand competition.} 
Seidmann (2009) considers a sequential model of an open-access rule with consent in a three-country setup, where countries can form bilateral FTAs, a bilateral CU, or the multilateral FTA. 
Utility transfer between countries is allowed, and the countries can renegotiate a PTA for expansion. These features lead to the result that efficient global FTA eventually forms. Seidmann's model shows a different motive (``strategic positioning'') of a PTA from ours that PTA members strategically manipulate the status quo, seeking advantages in future negotiations.

Missios et al.\ (2016), Furusawa and Konishi (2007), and Goyal and Joshi (2006) consider the formation of FTAs based on the stability approach. 
Missios et al.\ (2016) employ a coalitional-proof Nash equilibrium in a three-country setup, and Furusawa and Konishi (2007) and Goyal and Joshi (2006) employ the pairwise stability of network formation games in an $n$-country setup. 
All of them show that global FTA is a unique stable outcome if countries are symmetric.%
\footnote{Goyal and Joshi (2006) prove that a pairwise stable network is either global FTA or the second-largest FTA with a single non-participant in a Cournot competition model.} 
Unlike these works, our approach is process-based. 
We explicitly formulate a sequential trade liberalization process as a non-cooperative game and analyze a subgame perfect equilibrium. 
Two approaches are complementary. 
We discuss differences between our results and the stability results in the literature in Section~6.

The rest of the paper is organized as follows.
Section~2 presents the underlying economy. 
Section~3 considers the Nash equilibrium tariff in the non-cooperative tariff game. 
Section~4 investigates the formation of an FTA under two open-access rules. 
Section~5 illustrates a numerical example. 
Section~6 discusses the results, related to the literature. Section~7 concludes.
The proofs are given in Appendices and the supplementary materials, and tedious calculations are collected in the supplementary materials.\footnote{Available upon request.} 

\section{The Economy}

We consider a perfectly competitive economy with $n$ ``large'' countries indexed by $i=1, \cdots, n$.%
\footnote{We generalize the two-country model of Horn et al. (2010) to an $n$ country case. Missios et al. (2016) and Saggi et al. (2018) also consider multi-country versions of the model.} 
There are $n$ non-numeraire goods indexed by $J=1, \cdots, n$, and one numeraire good indexed by $0$. 
A consumption vector is represented by $c=(c^1, \cdots, c^n, c^0)$, where each $c^J$ $(J=1, \cdots, n)$ is the consumption of non-numeraire good $J$ and $c^0$ is that of the numeraire good. 

A representative consumer in every country $i$ has the following utility function, which is additively separable and linear in the numeraire good:
\begin{align*}
U_i(c)=\sum_{J=1}^n u_i(c^J) + c^0 , 
\end{align*} 
where $u_i(c^J)$ is $i$'s utility function for consumption $c^J$ of each non-numeraire good $J$. 
For every $J$, $u_i(c^J)$ is quadratic:
\begin{align}
u_i(c^J)= a c^J - \frac{1}{2} (c^J)^2.
\label{eq:2_1}
\end{align}
Let $p_i^J$ be the price of each good $J$ in country $i$. 
By \eqref{eq:2_1}, country $i$'s demand for good $J$ is given by
\begin{align}
d_i^J(p_i^J)= a - p_i^J.
\label{eq:2_2}
\end{align}

Labor is only the production factor. Every one unit of numeraire good is produced by one unit of labor. 
We assume that labor supply is large enough to ensure strictly positive production, and therefore the equilibrium wage is equal to one. 

Each non-numeraire good is produced from labor with diminishing returns. 
Country $i$'s production function for non-numeraire good $J$ is given by
\begin{align*}
Q_i^J=\sqrt{2\lambda_i^J l^J} , 
\end{align*}
where $Q_i^J$ is country $i$'s production of good $J$ and $l^J$ is labor input for the production of good $J$. 
Each country $i$ produces all $n$ goods $J=1, \cdots, n$.

By assuming an inner solution of the profit maximization,%
\footnote{The supply $s_i^J$ of good $J$ is determined by the labor input $l^J$ to maximize profit $p_i^J \sqrt{2\lambda_i^J l_J} - l^J$, where $s_i^J=\sqrt{2\lambda^J_il^J}$.} 
we obtain country $i$'s supply function of good $J$, given by
\begin{align}
s_i^J(p_i^J)= \lambda_i^J  p_i^J.
\label{eq:2_3}
\end{align}

With respect to the trade pattern in the world, each country $i$ has a comparative advantage in one good which is indexed by the upper letter $I$, and it has a comparative disadvantage in all other goods. 
In particular, we assume that for every $i$ and every good $J$ with $J \ne I$,
\begin{align}
\lambda_i^I=1 + \lambda, \hspace{1em}  \lambda_i^J=1 , 
\label{eq:2_4}
\end{align}
where $\lambda > 0$ is the degree of comparative advantage. 

Countries' comparative advantage structure implies that each country $i$ is a sole exporter of good $I$, and imports $n-1$ other goods $J \ne I$ from countries $j \ne i$. 
All countries except country $i$ compete with each other for importing good $I$ from country $i$. 
From this property, the economy is called the \emph{competing importers} model (Missios et al. 2016).%
\footnote{Saggi and Yildiz (2010) and Saggi et al. (2013, 2019) consider a ``competing exporters'' model where each country is a sole importer of a given good.}

Let $t_{ij}$ be a tariff imposed by country $i$ on its imports of good $J (\ne I)$ from country $j$. 
A vector $t=(t_{ij}: i, j=1, \cdots, n, i\ne j)$ is a tariff profile in the world. 
When consumers in country $j$ buy good $I$ imported from country $i$, they pay the price $p_i^I+ t_{ji}$, where $p_i^I$ is good $I$'s world price. 

Let $p_j^I$ be country $j$'s domestic price of good $I$. 
By ruling out prohibitive tariffs, we assume the no-arbitrage condition for good $I$ for every country $j (\ne i)$:
\begin{align}
p_j^I = p_i^I+ t_{ji} .
\label{eq:2_5}
\end{align}
Country $j$'s domestic price $p_j^I$ of good $I$ is equal to the import price $p_i^I$ plus the tariff $t_{ji}$.

The world market clearing condition on each good $I$ is given by
\begin{align}
\sum_{j=1}^n d_j^I(p_j^I)=\sum_{j=1}^n s_j^I(p_j^I) , 
\label{eq:2_6} 
\end{align}
where $d_j^I$ is country $j$'s demand of good $I$ and $s_j^I$ is country $j$'s supply (production) of good $I$.

Let $m_j^I$ be country $j$'s imports of good $I$ from country $i$, where
\begin{align}
m_j^I(p_j^I)=d_j^I(p_j^I)- s_j^I(p_j^I) , 
\label{eq:2_7}
\end{align}
and let $x_j^I$ be country $i$'s exports of good $I$ to country $j$, where
\begin{align}
x_j^I(p_j^I)=s_i^I(p_i^I)- d_i^I(p_i^I) - \sum_{k \ne i, j}m_k^I(p_k^I) . 
\label{eq:2_8}
\end{align}

From the market clearing condition \eqref{eq:2_6} for good $I$, country $i$'s exports to country $j$ are equal to country $j$'s imports from country $i$.
That is, 
\begin{equation}
x_j^I(p_j^I)=m_j^I(p_j^I) . 
\label{eq:2_9}
\end{equation}
Conversely, if \eqref{eq:2_9} holds for every country $j (\ne i)$, the world market clearing condition is satisfied. 

A \textit{market equilibrium} under a tariff profile $t$ is defined by a collection of prices $p=(p_{i}^J: i, J=1, \cdots, n)$ which satisfies the no-arbitrage condition \eqref{eq:2_5} and the world market clearing condition \eqref{eq:2_6} for every country $i$ and every good $J$. 
\vspace{2ex}

\noindent\textbf{Lemma~2.1.} The equilibrium prices under a tariff profile $t$ are given by
\begin{align}
p_i^I=\frac{na - 2\sum_{j \ne i} t_{ji}}{2n + \lambda}, \hspace{1em} 
p_j^I=\frac{na - 2\sum_{k \ne i, j} t_{ki} + (2n-2+\lambda)t_{ji}}{2n + \lambda} . 
\label{eq:2_10}
\end{align}

The lemma shows how a tariff profile of large countries affects the world and the domestic prices of every good. 
Ceteris paribus, the increase of country $j$'s tariff $t_{ji}$ on country $i(\ne j)$ has both the \emph{direct} and \emph{indirect} effects on prices. 
First, the tariff increase of $t_{ji}$ raises good $I$'s domestic price $p_j^I$ in country $j$ (the direct effect). 
Second, the tariff increase of $t_{ji}$ reduces good $I$'s world price $p_i^I$ and its domestic prices $p_k^I$ in all other countries $k(\ne i,j)$ (the indirect effect).

The above price changes caused by tariffs induce the two effects on the demand, supply, and imports of goods in the world.
First, the tariff increase of $t_{ji}$ reduces the demand $d_j^I$ of good $I$ in country $j$, raises its supply $s_j^I$, and reduces its import $m_j^I$ (the direct effect). 
Second, the tariff increase of $t_{ji}$ raises the demand $d_k^I$ of good $I$ in a third country $k(\ne i,j)$, reduces its supply $s_k^I$, and raises its imports $m_k^I$ (the indirect effect).

We remark that the indirect effect is rooted in the law of supply that \eqref{eq:2_3} is upward-sloping.\footnote{By contrast, the frameworks of Cournot oligopoly and competing exporters assume a perfectly elastic supply function.}
Since country $j$ attempts to keep its total supply by the law, the decrease of $j$'s exports to country $i$ is partially covered by the increase of $j$'s exports to a third country $k$. 


The price changes by tariffs derived from Lemma~2.1 further show how FTA formation affects the trade pattern in the world. 
If two countries $i$ and $j$ forms an FTA to eliminate tariffs between them, it causes the following three effects, ceteris paribus. 
First, the FTA member countries $i$ and $j$ raise their imports of good $I$ and $J$, due to the domestic price decrease. 
Second, a third country $k(\ne i,j)$ reduces its imports of the goods $I$ and $J$, due to the domestic price increase.
Third, FTA member countries' imports of good $K$ from country $k$ are unchanged because markets of different goods are independent in the model. 

If a third country $k$ can adjust its tariffs on FTA member countries, then the effect of the FTA formation becomes more complicated. 
We will explore this issue in the next section by analyzing a tariff game after an FTA is formed. 


Every country $i$'s welfare $W_i(t)$ under a tariff profile $t$ is defined as the sum of consumer surplus, producer surplus, and tariff revenue for all goods:
\begin{equation}
W_i(t)=\sum_{J=1}^n CS_i^J(p_i^J) + \sum_{J=1}^n PS_i^J(p_i^J) + \sum_{j \ne i} t_{ij}m_i^J(p_i^J) , 
\label{eq:2_11}
\end{equation}
where $CS_i^J$ is country $i$'s consumer surplus for good $J$ and $PS_i^J$ is its producer surplus for good $J$. 

From \eqref{eq:2_1} and \eqref{eq:2_2}, 
\begin{equation}
CS_i^J(p_i^J)=u_i(d_i^J(p_i^J))- p_i^J d_i^J(p_i^J)=\frac{1}{2}(a-p_i^J)^2 . 
\label{eq:2_12}
\end{equation}
From \eqref{eq:2_3}, 
\begin{equation}
PS_i^J(p_i^J)= \int_0^{p_i^J} s_i^J(p) d p = \frac{1}{2} \lambda_i^J  (p_i^J)^2 . 
\label{eq:2_13}
\end{equation}
By \eqref{eq:2_4}, \eqref{eq:2_11}--\eqref{eq:2_13}, $W_i(t)$ is calculated as
\begin{equation}
W_i(t)=\frac{1}{2}\sum_{J=1}^n (a-p_i^J)^2 + \frac{1}{2}\{(1+\lambda)(p_i^I)^2 + \sum_{J \ne I} (p_i^J)^2 \} + \sum_{j \ne i} t_{ij}(a-2p_i^J) . 
\label{eq:2_14}
\end{equation}

%

Finally, we remark that the model has the following simple structures: 
(i) the markets of different goods in every country are separated, and thus the economy is in a partial equilibrium set-up, 
(ii) every country exports a unique good, in which it has a comparative advantage, 
(iii) every country's tariff on one exporter never affects its imports from other exporters, 
(iv) the choice of a country's tariff on a given country is independent of its tariffs on other countries,
(v) all countries are symmetric in demand, supply and comparative advantage structure.%
\footnote{We further prohibit all countries from tariff deflection.}
Thanks to these properties, the economy provides a tractable model to study trade liberalization in an inter-industry framework.

\section{The Non-cooperative Tariff Game}

We consider the following $n$-country non-cooperative tariff game. 
All countries $i$ simultaneously choose their tariffs $t_i=(t_{ij}: j=1, \cdots, n, j\ne i)$ to maximize the welfare $W_i(t)$ 
given by \eqref{eq:2_11}. 
Given the tariff profile $t=(t_{i}: i=1, \cdots, n)$, the competitive equilibrium prices \eqref{eq:2_10} prevail. 
Differentiating $W_i(t)$ with respect to $t_{ij}$ yields
\begin{equation}
\frac{\partial W_i}{\partial t_{ij}} 
= \frac{\partial CS_i^J(p_i^J)}{\partial t_{ij}} + \frac{\partial PS_i^J(p_i^J)}{\partial t_{ij}} + m_i^J(p_i^J) + t_{ij} \frac{\partial m_i^J(p_i^J)}{\partial t_{ij}} , 
\label{eq:3_1}
\end{equation}
where $m_i^J(p_i^J)$ is country $i$'s imports of good $J$. 
From \eqref{eq:2_7}, \eqref{eq:2_12} and \eqref{eq:2_13}, \eqref{eq:3_1} is rewritten as%
\footnote{The derivation is given in the supplementary material S.1.}
\begin{equation}
\frac{\partial W_i}{\partial t_{ij}} = t_{ij} \frac{\partial m_i^J}{\partial p_i^J} \frac{\partial p_i^J}{\partial t_{ij}} - m_i^J \frac{\partial p_j^J}{\partial t_{ij}} . 
\label{eq:3_2}
\end{equation}

Similarly to the effects on equilibrium prices mentioned in the last section, the tariff change has the following two effects on the welfare. 
We can interpret the first (negative) term of \eqref{eq:3_2} as the efficiency cost of the tariff. 
It gives country $i$'s marginal welfare loss from its own tariff $t_{ij}$ on good $J$. 
The marginal increase of the domestic price $p_i^J$ of good $J$, or \emph{the pass through} of the tariff, reduces the imports. 
The second (positive) term of \eqref{eq:3_2} can be interpreted as the effect of terms of trade. 
Since country $i$ imports good $J$, $i$'s welfare is affected by the marginal decrease of the world price $p_j^J$ of good $J$, or \emph{the terms of trade gain} of the tariff, due to $i$'s tariff $t_{ij}$ on country $j$. 

From \eqref{eq:2_14},
\begin{equation}
\frac{\partial W_i}{\partial t_{ij}} =  -2t_{ij} + \frac{2}{2n+\lambda}(a-2p_j^J) , 
\label{eq:3_3}
\end{equation}
by using $\frac{\partial p_i^J}{\partial t_{ij}}=\frac{2n-2+\lambda}{2n+\lambda}$
(see \eqref{eq:2_10}). 
The derivation of \eqref{eq:3_3} is given in the proof of Theorem~3.1. 
We note that the second term of \eqref{eq:3_3} is positive. 
If tariff $t_{ij}$ is low, a unilateral increase of $t_{ij}$ benefits country $i$. 
Otherwise, it may be harmful. 

By assuming the first-order condition, for every $j(\ne i)$, 
\eqref{eq:3_3} implies that the optimal tariff $t_{ij}$ of country $i$ is given by 
\begin{align}
t_{ij} = \frac{1}{2n+\lambda}(a-2p_j^J)  . 
\label{eq:3_4}
\end{align}

We observe from \eqref{eq:3_4} that the optimal tariff $t_{ij}$ on good $J$ is independent of $i$, i.e.\ the tariff $t_{ij}$ is identical across importing countries $i$ of good $J$. 
This property is derived from the symmetry assumption of the economy. 
In what follows, we will show the same kind of tariff symmetry when an FTA forms. 
For example, member countries in an FTA impose an identical external tariff on non-members.
\eqref{eq:3_4} also implies that every country $i$'s optimal tariff $t_{ij}$ on good $J$ is decreasing in the world price $p_j^J$. 

Substituting \eqref{eq:2_10} into \eqref{eq:3_4} yields country $i$'s best response function, 
which assigns $i$'s optimal tariff $t_{ij}$ on $j$ to all other countries' tariffs $(t_{kj})_{k\ne i,j}$, as follows. 
\vspace{2ex}
\begin{equation}
t_{ij}= \frac{a \lambda + 4 \sum_{k \ne i, j}t_{kj} }{ (2n+\lambda)^2-4 } .
\label{eq:3_5}
\end{equation}

The best response function \eqref{eq:3_5} shows that tariffs are \emph{strategic complements}. If other countries raise their tariffs on country $j$, country $i$ also raises its tariff on country $j$. 
The complementarity is caused by the fact that country $i$'s marginal welfare with respect to $i$'s tariff is increasing in country $k$'s tariff.\footnote{Formally, the strategic complementarity of tariffs is derived by $\frac{\partial^2 W_i}{\partial t_{ij}^2}<0$ and $\frac{\partial^2 W_i}{\partial t_{ij} \partial t_{kj}}>0$.}

We now obtain the following theorem.
\vspace{2ex}

\noindent \textbf{Theorem 3.1.} There exists a unique Nash equilibrium in the non-cooperative tariff game. 
All countries impose the common tariff $t^{\mathrm{NE}}$ on the imports of every good $J$, 
\begin{align}
    t^{\mathrm{NE}} = t_{ij} = \frac{a\lambda}{\lambda^2+4n\lambda+4(n^2-n+1)} .
\label{eq:3_6}
\end{align}

The Nash equilibrium of the tariff game describes the ``tariff war'' where all countries impose (strictly) positive tariffs on each other to maximize their individual welfare. 
Country $i$'s positive tariff $t_{ij}$ on the imports of good $J$ raises the domestic price $p_{i}^J$ of good $J$. 
This price increase reduces the consumer surplus and the imports of good $J$, and raises the producer surplus. 

While the effect on the tax revenue is ambiguous due to the import reduction, 
the best response \eqref{eq:3_5} shows that country $i$ optimally imposes a positive tariff, regardless of the other countries' tariffs. 
It is worth noting that the Nash equilibrium tariff $t^{\mathrm{NE}}$ converges to zero as the number of countries goes to infinity.

To conclude this section, we show that the Nash equilibrium is Pareto inefficient and that the global free trade is efficient to maximize the total (world) welfare. 

From \eqref{eq:2_11}, the total welfare $W$ is given by 
\begin{align*}
W &=\sum_{i=1}^n W_i \notag \\ 
  &=\sum_{i=1}^n\sum_{J=1}^n \left[ CS_i^J(p_i^J) + PS_i^J(p_i^J) \right] 
                  + \sum_{i=1}^n \sum_{j \ne i}^n t_{ij}m_i^J(p_i^J).
\end{align*}
To maximize the total welfare $W$, every country $i$ takes into account how its tariff $t_{ij}$ on good $J$ affects not only its own welfare $W_i$, but also all other countries' welfare. 

Similarly to \eqref{eq:3_3}, \eqref{eq:2_10} and \eqref{eq:2_14} give the effects of country $i$'s tariff $t_{ij}$ on others' welfare, given by
\begin{align}
\frac{\partial W_j}{\partial t_{ij}} &=  \frac{2}{2n+\lambda}(a-(2+\lambda)p_j^J) \label{eq:3_7} \\ 
\frac{\partial W_k}{\partial t_{ij}} &= \frac{2}{2n+\lambda}(a-2p_j^J)   \label{eq:3_8}
\end{align}
for every $k (\ne j)$.%
\footnote{The derivation is given in the supplementary material S.2.}
If country $i$'s comparative advantage degree $\lambda$ is low, an increase of country $i$'s tariff $t_{ij}$ on country $j$ may benefit country $j$, owing to the producer surplus increase.  
The increase of $t_{ij}$ always benefits a third country $k$ (i.e.\ $\frac{\partial W_k}{\partial t_{ij}} >0$).

The effect of country $i$'s tariff $t_{ij}$ on the total welfare $W$ is given by
\begin{align}
\frac{\partial W}{\partial t_{ij}}
&= \sum_{k=1}^n \frac{\partial W_k}{\partial t_{ij}} \notag \\
&= -2t_{ij} + \frac{2}{2n+\lambda} [na-(\lambda+2n)p_j^J] . 
\label{eq:3_9}
\end{align}
Substituting \eqref{eq:2_10} into \eqref{eq:3_9} yields
the first-order condition for country $i$'s optimal tariff on good $J$ to maximize the total welfare, which is given by
\begin{equation}
t_{ij} = \frac{2}{2n+\lambda}\sum_{k \ne j}t_{kj} . 
\label{eq:3_10}
\end{equation}

We can easily show that \eqref{eq:3_10} has a unique solution $t_{ij}=0$ for every $i$ and $j \ne i$. 
Thus, we have the following theorem.%
\footnote{In the supplementary material S.3, we prove part (i) without assuming the first-order condition.}

\vspace{2ex}
\noindent \textbf{Theorem 3.2.} 
\begin{enumerate}[(i)]
    \item Global FTA uniquely maximizes the total world welfare.
    \item Every country $i$'s welfare under global FTA is 
\begin{equation*}
\frac{na^2(n+\lambda)}{2(2n+\lambda)}.
\end{equation*} 
    \item Every country $i$'s Nash equilibrium welfare under no FTA is
\begin{equation*}
\frac{na^2(n+\lambda)}{2(2n+\lambda)} - \frac{(n-1)(2+\lambda)}{2n+\lambda} (t^{\mathrm{NE}})^2 , 
\end{equation*}
where $t^{\mathrm{NE}}$ is the Nash equilibrium tariff in \eqref{eq:3_6}.
\end{enumerate}
The theorem shows that our model satisfies the standard property of the literature on international trade, namely, global free trade is the efficient regime. 
Since global free trade is not a Nash equilibrium, each country has an incentive to raise its tariffs unilaterally. 
Such a self-interested behavior, however, harms other countries' welfare.

By Theorem~3.2, we observe how the model parameters affect the welfare of countries. 
Each country's welfare under global FTA (and under no FTA) increases as each of the number $n$ of countries, the demand $a$ and the degree $\lambda$ of comparative advantage increases. 

We also observe that every country's welfare is monotonically decreasing in a common tariff of all countries (see \eqref{Aeq:3_2} in Appendix~A.3). 
Therefore, it is optimal for every country to agree to global FTA \emph{under the constraint that all countries in the world establish identical tariffs}.
This fact theoretically supports the WTO's traditional approach to achieving global free trade under the most favored nation (MFN) principle and the reciprocity (Baldwin, 2016).

\section{Free Trade Agreements}

We now investigate the endogenous formation of an FTA whose members are bound to set zero tariffs on each other. The members impose independently their external tariffs on non-members. 

We present a sequential game where an FTA may emerge and grow. In the game, a non-member country can access freely to an incumbent FTA for participation. 
By this property, we call the process an \emph{open-access rule}.\footnote{Seidmann (2009) calls an open-access rule with consent in our terminology a closed-access rule, and calls an open-access rule without consent an open-access rule.}
The game proceeds with the following three stages.
\begin{enumerate}
\item All $n$ countries decide sequentially whether they participate in an FTA, according to a fixed order of moves. 
Without loss of generality, let the order be $(1, \cdots, n)$.

\item If a new country decides to participate in an FTA, then all incumbent members either accept or reject its participation simultaneously. 
The participation is approved if all incumbents unanimously accept it.%
\footnote{Since countries are symmetric in our model, the result is not affected critically if unanimity rule is replaced with majority rules.}  
When there is no incumbent, any participant forms an FTA unilaterally, and succeeding countries decide to participate in it, or not.

\item After a new FTA forms, all inside and outside countries choose their external tariffs independently. The internal tariff of the FTA is zero. 
\end{enumerate}

An important feature of our FTA formation game is that a new participation needs incumbent members' unanimous consent. 
We call this property an \emph{open-access rule with consent}. The game aims to model the formation process of open regionalism FTAs including APEC, as we have discussed in Introduction. 
Whenever each country moves, it knows perfectly all the previous other countries' choices.  

We will characterize a subgame perfect equilibrium (SPE) of the game to answer the following questions: 
(i) how the formation of FTAs affects tariff setting of member and non-member countries, 
(ii) how FTAs affect the welfare of member and non-member countries. 

We first examine countries' tariff setting behavior under the formation of FTAs, which may be overlapped. 
Suppose that several FTAs, $F_1, \cdots, F_m$, are formed, where each $F_i$ is a subset of countries. 
All member countries of each FTA impose zero internal tariffs. 
Neither of two FTAs is a subset of the other.\footnote{If $F_i$ is a subset of $F_j$, then the free trade agreement of $F_j$ includes that of $F_i$, and thus $F_i$ is redundant.} 
Two different FTAs can have common members, i.e.\ countries may participate in more than one FTA. 

We call a collection $(F_1, \cdots, F_m)$ an \textit{FTA regime} and denote it by $\mathcal{F}$. 
The non-cooperative tariff game in the last section where no country participate in an FTA corresponds to the FTA regime $(\{1\}, \cdots, \{n\})$.

For every country $k$, let $N_k$ be the set of countries with which country $k$ does not have any FTA, and let $n_k \ (0\leq n_k\leq n-1)$ be the number of countries in $N_k$. As $n_k$ increases, country $k$ participates in less FTAs. 
All countries $i$ in $N_k$ choose independently their tariffs on country $k$ to maximize their own welfare. 
We denote country $i$'s external tariff on country $k$ under an FTA regime $\mathcal{F}$ by $t_{ik}^{\mathcal{F}}$.
The number $n_k$, which determines the equilibrium tariff $t_{ik}^{\mathcal{F}}$ in Theorem~4.1, plays a key role. 

The first-order condition \eqref{eq:3_4} of the welfare maximization of country $i\in N_k$ shows that $i$'s tariff $t_{ik}^{\mathcal{F}}$ on country $k$ is independent of $i$. 
Then, from \eqref{eq:3_5},
\begin{align}
 t_{ik}^{\mathcal{F}} = \frac{1}{ (2n+\lambda)^2-4}
 \left[ a \lambda + 4 (n_k-1)t_{ik}^{\mathcal{F}} \right]. 
\label{eq:4_1}
\end{align}
\eqref{eq:4_1} holds true because all countries $i$ in $N_k$ impose the identical external tariff $t_{ik}^{\mathcal{F}}$ on country $k$, and all countries outside $N_k$ impose zero tariff on country $k$.  

Solving \eqref{eq:4_1}, we obtain the following result.

\vspace{2ex}
\noindent \textbf{Theorem 4.1.} 
Let $\mathcal{F}=(F_1, \cdots, F_m)$ be an FTA regime, and for every $k=1,\cdots,n$, let $N_k$ be the set of countries with which country $k$ does not have any FTA. 
Then, every country $i$ in $N_k$ imposes the external tariff $t_{ik}^{\mathcal{F}}$ on country $k$ such that
\begin{align*}
  t_{ik}^{\mathcal{F}} = \frac{a\lambda}{\lambda^2+4n\lambda+4(n^2-n_k)} , 
\end{align*}
where $n_k$ is the number of countries in  $N_k$.%
\footnote{Saggi et al.\ (2018, Appendix~A) derive FTA non-member countries' external tariffs on member countries when a single FTA is formed.} 
\vspace{2ex}

\noindent The theorem provides an answer to our question (i) of how the formation of FTAs affects the tariff setting behavior of countries. 
Given the model parameters $(n,a,\lambda)$, country $i$'s external tariff $t_{ik}^{\mathcal{F}}$ on each country $k$ is solely determined by $n_k$, the number of countries with which country $k$ does not have FTAs. 

The external tariff $t_{ik}^{\mathcal{F}}$ is monotonically increasing in $n_k$. 
This property reveals the following two important effects of FTAs on the tariff setting of member and non-member countries. 
First, all FTA member countries impose the same external tariffs on non-member country $k$ as before its formation, provided that $n_k$ does not change. 
Thus, the model satisfies the Article XXIV of GATT to prohibit PTAs from increasing their external tariffs on non-member countries. 
Second, suppose that an FTA is formed. 
Then, non-member countries $k$ reduce their external tariffs $t_{ki}^{\mathcal{F}}$ on member countries because the number $n_i$ decreases owing to the FTA formation. 
This \emph{tariff complementarity} is implied by the best response function \eqref{eq:3_5} in the tariff game. 

Theorem~4.1 also shows that the external tariff on non-members is independent of an FTA size. 
If country $k$ participates in no FTA ($n_k=n-1$), the external tariff $t_{ik}^{\mathcal{F}}$ on country $k$ is equal to the Nash equilibrium tariff $t_k^{\mathrm{NE}}$ in the non-cooperative tariff game.  
The theorem implies that the Nash equilibrium tariff is the highest one among all tariffs under FTA regimes.

When an FTA is formed, internal tariffs of the FTA are set to be zero, and non-member countries adjust their external tariffs on the FTA.
The next theorem shows how those tariff changes of the FTA formation affect the trade pattern in the world.

\vspace{2ex}
\noindent \textbf{Theorem 4.2.} 
Let $\mathcal{F}^0=(\{1\}, \cdots, \{n\})$ and $\mathcal{F}^i=(F,\{m+1\},\cdots,\{n\})$, where $m$ is the size of an FTA $F$ and $i\in F$. 
Let $p_j^{I,\mathcal{F}}$ and $m_j^{I,\mathcal{F}}$ be the price and imports of good $I$ in country $j$ under an FTA regime $\mathcal{F}\in\{\mathcal{F}^0,\mathcal{F}^i\}$, respectively.
Then,
\begin{enumerate}[(i)]
    \item $p_j^{I,\mathcal{F}^0} > p_j^{I,\mathcal{F}^i}$ and $m_j^{I,\mathcal{F}^0} < m_j^{I,\mathcal{F}^i}$ for every member $j\in F$ ($j\neq i$), 
    \item $p_k^{I,\mathcal{F}^0} < p_k^{I,\mathcal{F}^i}$ and $m_k^{I,\mathcal{F}^0} > m_k^{I,\mathcal{F}^i}$ for every non-member $k\not\in F$.
\end{enumerate}
\vspace{2ex}
Theorem~4.2 shows that an FTA $F$ with a member country $i$ reduces the domestic prices of good $I$ in other member countries $j$ from country $i$, and raises their imports of good $I$. 
The FTA formation has the opposite effect on non-member countries $k$, i.e.\ it raises their domestic prices of good $I$ and reduces their imports.

As the equilibrium price \eqref{eq:2_10} shows, the domestic price of the imported goods from an FTA member country in a non-member country is affected not only by its own tariff but also by other non-member countries' tariffs. 
It is increasing in the former and decreasing in the latter.

All non-member countries' tariffs on the FTA member countries are identical due to the symmetry assumption. 
It turns out that the marginal increase of the domestic price with respect to non-members' tariff is equal to $\frac{2m+\lambda}{2n+\lambda}$, where $m$ is the size of the FTA (see \eqref{Aeq:4_4}). 
Non-member countries' tariff is decreasing in the FTA size $m$ due to the strategic complementarity shown by the best response function \eqref{eq:3_5}. 

Combining these two effects, we observe that the domestic price of the imported goods in a non-member country is increasing in the FTA size. 
In particular, the FTA formation $(m>1)$ induces its domestic price increase of goods imported from member countries, compared to the case of no FTA ($m=1)$. 

The domestic price increase reduces non-member countries' imports from an FTA. 
Saggi et al.\ (2018) call this phenomenon the \emph{external trade diversion}. 
They empirically support this theoretical prediction, using industry-level data on all FTAs formed in the world during 1989-2011.

We now proceed to our question (ii) of how FTAs affect the welfare of both member and non-member countries. 
To answer the question, we will evaluate every country's welfare under all possible FTA regimes. 
In the following lemma, we first derive an explicit formula of every country's welfare under any tariff profile.\footnote{Since the derivation of \eqref{eq:4_2} is tedious, we prove the lemma in the supplementary material S.4.}
\vspace{2ex}

\noindent \textbf{Lemma 4.1.}  
Let $r=2(2n+\lambda)^2$. 
Every country $i$'s welfare $W_i(t)$ under a tariff profile $t=(t_{ij}: i, j=1, \cdots, n, i\neq j)$ is given by 
\begin{align}
 rW_i(t) 
=& na^2(n+\lambda)(2n+\lambda) \notag \\
&+ \sum_{j\ne i} t_{ij} \left[ 4a\lambda+16\sum_{k\ne i,j}t_{kj} \right] - \sum_{j\ne i} (t_{ij})^2 \left[ 2 (2n+\lambda)^2-8 \right] \notag \\ 
&+ 4(2+\lambda)\left( \sum_{k\ne i}t_{ki} \right)^2 - 4a\lambda(n-1) \sum_{k\ne i}t_{ki} \notag \\
&+ 4a\lambda \sum_{j\ne i} (\sum_{k\ne i,j} t_{kj}) + 8 \sum_{j\ne i} (\sum_{k\ne i,j} t_{kj})^2.
\label{eq:4_2}
\end{align}

Employing \eqref{eq:2_10} in Lemma~2.1, we can compute every country's welfare under an FTA regime $\mathcal{F}=(F_1, \cdots, F_m)$, denoted by $W_i^\mathcal{F}(t)$, where $t$ is the equilibrium tariff profile under $\mathcal{F}$. 
All countries $j$ in $N_i$, who have no FTAs with country $i$, impose positive tariffs $t_{ji}$ on country $i$.
From Theorem 4.1, the tariff $t_{ji}$ is solely determined by the number $n_i$, which is the size of $N_i$, and thus it is independent of index $j$. 
In the following, we denote $t_{ji}$ by $t_{i}$, whenever no confusion arises.
Then, \eqref{eq:4_2} is rewritten as 
\begin{align}
rW_i^{\mathcal{F}}(t) 
=& na^2(n+\lambda)(2n+\lambda) + \sum_{j\in N_i} \left[4a\lambda+16(n_j-1)t_{j}\right]t_j \notag \\
&-\left[ 2 (2n+\lambda)^2-8 \right]\sum_{j \in N_i} (t_j)^2 + 4(2+\lambda)\left( n_i t_{i} \right)^2 - 4a\lambda(n-1) n_it_{i} \notag \\
&+4a\lambda \sum_{j\in N_i} (n_j-1) t_{j}+4a\lambda \sum_{j\notin N_i, j\neq i} n_j t_{j} \notag \\
&+8 \sum_{j\in N_i} ((n_j-1) t_{j})^2 + 8 \sum_{j\notin N_i, j\neq i} (n_j t_{j})^2 . 
\label{eq:4_3}
\end{align}

\eqref{eq:4_3} reveals how each country's welfare depends on a FTA network. 
It depends not only on the number of countries with which a country is connected via FTAs, but also on the number of countries with which its partner countries are further connected. 
We will detail every country's welfare to investigate the endogenous FTA formation, which is our central issue.

By the game's rule, at most a single FTA can form. 
Without loss of generality, we denote an FTA by $F=\{1, \cdots, m\}$, where $1\leq m \leq n$. 
Thus, a possible FTA regime is given by $\mathcal{F}^m=(F, \{m+1\}, \cdots, \{n\})$. 
For every member $i \in F$, $n_i=n-m$, and for every non-member $j$, $n_j=n-1$.

As shown in Theorem 4.1, country $i$'s external tariff $t_{ij}$ against country $j$ is independent of $i$. 
In what follows, we denote the external tariff against every participant ($\mathit{p}\in F$) in the FTA by $t_{p}$, and the external tariff against every non-participant ($\mathit{np}\notin F$) in the FTA by $t_{np}$. 
By Theorem 4.1, under an FTA regime $\mathcal{F}^m =(F, \{m+1\}, \cdots, \{n\})$, the external tariff on every participant of the FTA is given by
\begin{align*}
t_{p} &= \frac{a\lambda}{\lambda^2+4n\lambda+4(n^2-n+m)} , 
\intertext{ and the external tariff on every non-participant is given by }
t_{np} &= \frac{a\lambda}{\lambda^2+4n\lambda+4(n^2-n+1)} .
\end{align*}

We also denote every participant's welfare by $W_{p}^m(t)$, and every non-participant's welfare by $W_{np}^m(t)$ under the FTA regime $\mathcal{F}^m$.

\vspace{2ex}
\noindent \textbf{Lemma 4.2.}  
Let $r=2(2n+\lambda)^2$ and $m$ be the number of FTA participants. 
Every participant's welfare $W_{p}^m(t)$ is given by 
\begin{align}
rW_{p}^m(t) =& na^2(n+\lambda)(2n+\lambda) + 4(2m+\lambda)(n-m)^2 t_{p}^2 \notag \\
& - 4a\lambda(n-m)^2t_{p} \notag \\
& + 2(n-m)\left[4-8n-4n\lambda - \lambda^2 \right]t_{np}^2 \notag \\
& + 4a\lambda (n-m)(n-1)t_{np} . 
\label{eq:4_4}
\end{align}
Every non-participant's welfare $W_{np}^m(t)$ is given by 
\begin{align}
rW_{np}^m(t) =& na^2(n+\lambda)(2n+\lambda) \notag \\
& -2m\left[\lambda^2+4n\lambda+8mn-4m^2\right]t_{p}^2 \notag \\
& -2\left[(n-m-1)\lambda^2-2(2mn-n^2+1)\lambda+4(n^2-2mn-n+m)\right]t_{np}^2 \notag \\
& +4a\lambda m(n-m)t_{p}-4a\lambda m(n-1)t_{np} . 
\label{eq:4_5}
\end{align}

\noindent The lemma follows from Lemma 4.1. 
Since the derivation is tedious, we prove the lemma in the supplementary material S.5. 
In what follows, we omit the equilibrium tariff profile $t$ in notations $W_{p}^m(t)$ and $W_{np}^m(t)$, and simply write $W_{p}^m$ and $W_{np}^m$ whenever no confusion arises.

The next two theorems show how FTA formation affects countries' welfare.

\vspace{2ex}
\noindent\textbf{Theorem 4.3.} 
$W_{p}^{m+1}>W_{np}^m$ for every $m=1, \cdots, n-1$.
\vspace{2ex}

The theorem shows that every non-participant's welfare improves if it joins an FTA. 
Thus, every non-participant has an incentive to join an FTA. 
We can explain its intuition as follows. 
The FTA growth removes all incumbent participants' tariffs on a new participant. 
The tariff complementarity, shown in Theorem~4.1, reduces non-participant's tariffs on a new participant.
Thus, a new participant can import at lower prices from all other countries, and then its welfare improves. 

By contrast, the FTA growth has mixed effects on incumbent participants' welfare. 
First, since incumbents' tariffs on non-participants and their imports from non-participants are constant, the FTA growth does not change their welfare yielded from non-participants.
Second, incumbents' tariff elimination on a new participant reduces their tariff revenue from it. 
Since the incumbents' imports from a new participant increase due to the direct effect of the tariff elimination, shown in Lemma~2.1, their consumer surplus yielded from a new participant rises. 
On the other hand, by their domestic price reduction due to the direct effect, the tariff elimination reduces their domestic supplies and producer surplus from a new participant.

Third, the tariff elimination by the FTA reduces each incumbent's consumer surplus and raises producer surplus yielded from other incumbents. 
Since the tariff complementarity decreases non-participants' tariffs on incumbents, its indirect effect reduces the imports among incumbents and raises the domestic supplies. 
The sum of these mixed effects by the FTA growth either raises or reduces incumbent participants' welfare.

The next theorem, however, shows that an FTA harms non-participants' welfare.

\vspace{2ex}
\noindent\textbf{Theorem 4.4.} 
$W_{np}^m$ is monotonically decreasing in FTA size $m$.
\vspace{2ex}

By the theorem, FTA formation has a negative externality to non-participants.
The external trade diversion can explain its intuition.
As shown in Lemma~2.1, the indirect effect of incumbent participants' tariff elimination reduces non-participants' imports from a new participant. 
Since non-participants produce those goods with their disadvantageous technology, the FTA growth reduces their consumer surplus.
Furthermore, since the tariff complementarity reduces non-participants' external tariffs on participants, their tariff revenues also reduce. 

By contrast, since all countries' external tariffs on a non-participant are unchanged, non-participants do not change their exports to all countries. 
Thus, the FTA growth does not change their producer surplus. 
In total, the FTA growth reduces every non-participant's welfare. 

We are now ready to characterize an SPE under the open-access rule with consent. For every $i=1, \cdots, n$, we call the stage game where country $i$ decides to participate in the FTA the \textit{$i$-th stage game}. 

The open-access rule with consent has the unanimous voting stage by incumbents. 
It is well-known that the simultaneous-move unanimous voting game has multiple Nash equilibria, which may involve weakly dominated strategies for voters. For example, the situation that all members reject new participation is a trivial Nash equilibrium of the unanimous voting game, while ``No'' vote may be weakly dominated by ``YES'' vote for each member.
To avoid this multiplicity, we restrict the analysis to an SPE with no weakly dominated strategies.\footnote{An alternative way to avoid the multiple Nash equilibria in the unanimity game is to employ a sequential-move game rather than a simultaneous-move game.}
\vspace{2ex}

\noindent\textbf{Theorem 4.5.} 
Suppose that the each participant's welfare $W_{p}^m$ in an FTA of size $m$ is maximized at $m^* \in \{2,\cdots, n\}$. 
Then, the open-access rule with consent has a unique SPE outcome, where only the first $m^*$ countries participate in an FTA.
\vspace{2ex}

The theorem shows that the open-access rule with consent does not lead to the global free trade unless it maximizes every participant's welfare. 
If the FTA size exceeds the optimal level of the participants' welfare, it is beneficial for them to reject new participantion because their welfare decrease, otherwise. 
If the FTA size is below the optimum, the incumbents accept a new participant, anticipating rationally that the optimal FTA size will attain in the subsequent process.

The following theorem shows that the equilibrium FTA size $m^*$ is strictly less than the global free trade, and is roughly $n/2$ for sufficiently large $n$. 

\vspace{2ex}

\noindent\textbf{Theorem~4.6.} 
The equilibrium FTA size $m^*$ under the open-access rule with consent is strictly less than $n$.
Furthermore, if $n$ is sufficiently large, 
$m^*$ is in $\{\frac{n}{2}-1$, $\frac{n}{2}, \frac{n}{2}+1\}$ for even $n$, and  
in $\{\frac{n-1}{2}, \frac{n+1}{2}\}$ for odd $n$. 
\vspace{2ex}

The unique SPE under the open-access rule with consent is consistent with our observation in the current international economy that open regionalism FTAs like APEC do not lead to global FTA. Complicated negotiations take place between new applicants and incumbent members, and new  participation has not been accepted. 

To implement open regionalism, we propose an open-access rule to accept any participant without consent of incumbent members. This new rule eliminates the unanimous voting stage in the FTA formation game.  
We call it an \emph{open-access rule without consent}. 
The open-access rule without consent is along the spirit of GATT Article I of General Most-Favoured-Nation Treatment.

We finally obtain the following result.

\vspace{2ex}
\noindent\textbf{Theorem 4.7.} 
The global free trade is a unique SPE outcome under the open-access rule without consent. 
The SPE outcome is independent of an order of moves. 
\vspace{2ex}

The theorem is derived from the fact that every non-participant is willing to join in an FTA, regardless of its size (Theorem 4.3). 
The unique SPE captures a preferable process that an FTA emerges and grows to the global free trade in  sequential negotiations. 
The theorem provides a theoretical support for the policy recommendation that the WTO should require any PTA to be open-access without incumbents' consent.

\section{A Numerical Example}

We provide a numerical example of the result when $n=8$, $a=12$ and $\lambda=36$. 
Table~1 gives the values of seven variables,%
\footnote{Notice that all the values are rounded to no more than six significant figures.}
(i) tariff $t_{p}$ on a participant of an FTA imposed by a non-participant, (ii) tariff $t_{np}$ on a non-participant imposed by a participant and another non-participant, (iii) participant's import $m_{p}^{p}$ from a participant in an FTA, (iv) non-participant's import $m_{np}^{p}$ from a participant, 
(v) participant's import $m_{p}^{np}$ from a non-participant,
(vi) participant's welfare $W_p$, and (vii) non-participant's welfare $W_{np}$, for each FTA size $m=1, \cdots, 8$. Non-participant's import $m^{np}_{np}$ from a non-participant is equal to $m^{np}_p$.
Figure~1 depicts the welfare of a participant and of a non-participant.

\begin{table}
    \centering \small
    \begin{filecontents*}{num_values1.csv}
$m$,$t_p$,$t_{np}$,$m_p^{p}$,$m_{np}^{p}$,$m_p^{np}=m_{np}^{np}$,$W_{p}$,$W_{np}$ 
1,$-$,0.161435,$-$,$-$,8.07175,$-$,487.251
2,0.161194,0.161435,8.38209,8.0597,8.07175,487.58,487.146
3,0.160954,0.161435,8.3696,8.04769,8.07175,487.804,486.937
4,0.160714,0.161435,8.35714,8.03571,8.07175,487.924,486.624
5,0.160475,0.161435,8.34473,8.02377,8.07175,487.942,486.209
6,0.160237,0.161435,8.33234,8.01187,8.07175,487.857,485.691
7,0.16,0.161435,8.32,8,8.07175,487.671,485.072
8,$-$,$-$,8.30769,$-$,$-$,487.385,$-$ 
\end{filecontents*}
\csvreader[tabular=c|ccccccc,
        head=false,
        table head=\toprule,
        late after line=\\,
        late after first line=\\\midrule,
        table foot=\bottomrule,
        ]{num_values1.csv}{}{\csvcoli & \csvcolii & \csvcoliii & \csvcoliv & \csvcolv & \csvcolvi & \csvcolvii & \csvcolviii}
    \caption{The effect of an FTA size at parameters $(n,a,\lambda)=(8,12,36)$. Notations: $m$ (FTA size), $t_{p}$ (tariff on a participant), $t_{np}$ (tariff on a non-participant),
 $m_{p}^{p}$ (participant's import from a participant), $m_{np}^{p}$ (non-participant's import from a participant), $m^{np}_p$ (participant's import from a non-participant), $W_p$ (participant's welfare), $W_{np}$ (non-participant's welfare).}
    \label{fig:numerical_values}
\end{table}

We observe the following from Table 1: 
(1) the external tariff on participants in the FTA decreases as the FTA size grows, 
(2) the external tariff on a non-participant is constant on the FTA size, 
(3) the trade between two participants decreases as the FTA grows, 
(4) non-participant's imports from the FTA decreases as the FTA grows, 
(5) if a non-participant joins an FTA, then its imports from a participant increase, 
(6) non-participant's exports to both a participant and a non-participant are constant in the FTA size, 
(7) the participant's welfare is maximized at $m^*=5$, 
(8) non-participant's welfare is monotonically decreasing in the FTA size, and 
(9) a non-participant is better-off by joining the FTA. 
Those observations are consistent with our results.

Our main result shows that the largest FTA (global free trade) forms under the open-access rule without consent, while only five countries form an FTA under the open-access rule with consent. 
Under the global FTA, the welfare of every country is $487.385$. 
Under the open-access rule with consent, each participant of the equilibrium FTA obtains the highest welfare $487.942$, but the welfare of each non-participant is $486.209$.

\begin{figure}
    \centering
    \includegraphics[width=.5\textwidth]{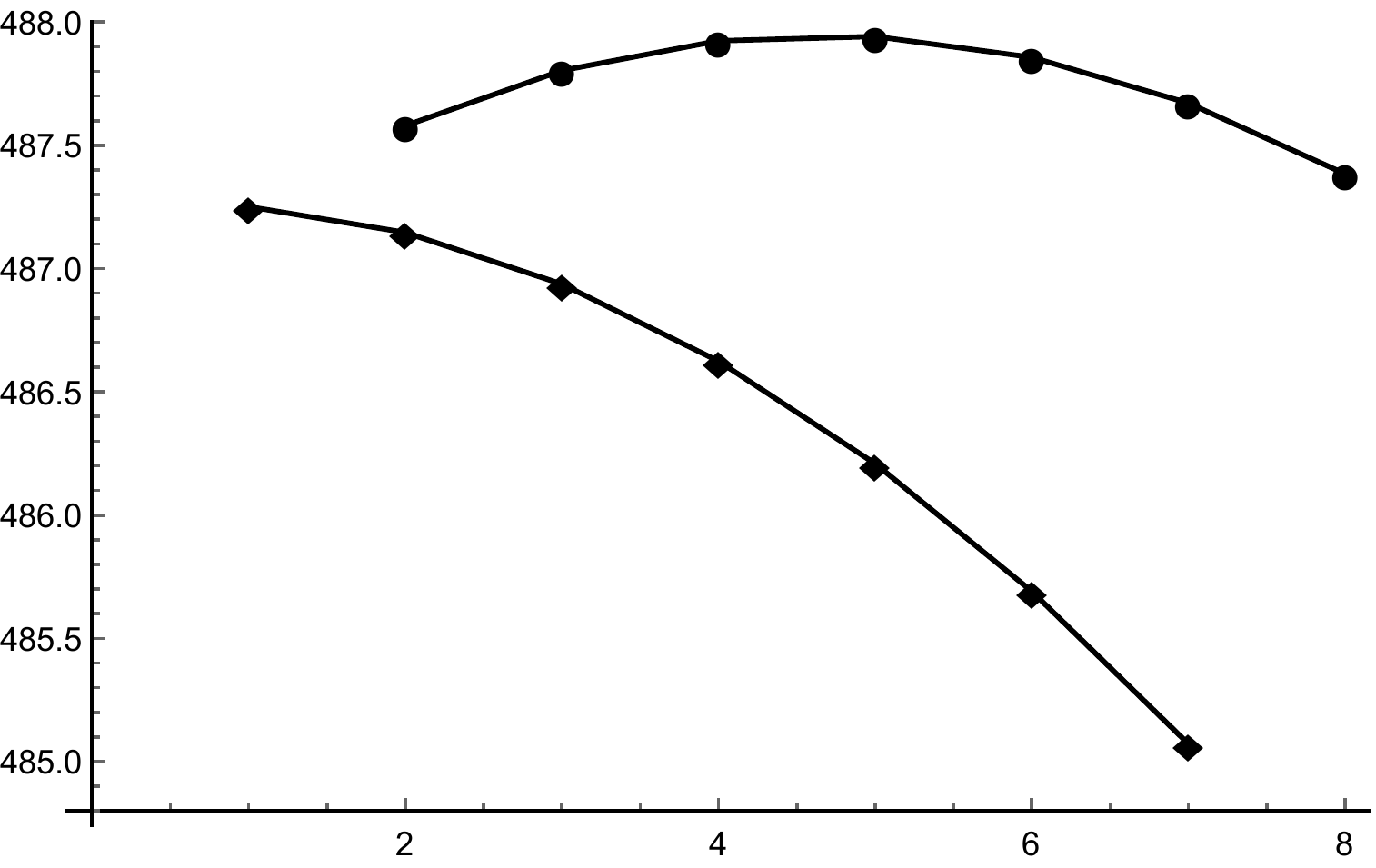}
    \caption{The welfare of a country. Note: The circle symbol depicts the welfare of a participant and the diamond symbol depicts the welfare of a non-participant. The $x$-axis designates an FTA size $m$.}
    \label{fig:numerival_graph}
\end{figure}

In the numerical example, first five countries form an FTA under the open-access rule with consent. 
For example, if the order of moves is $(1, 2, 3, 4, 5, 6, 7, 8)$, the FTA regime is $\mathcal{F}=(\{1, 2, 3, 4, 5\}, \{6\}, \{7\}, \{8\})$ in a unique SPE. 
Participation of the last three countries $6, 7$ and $8$ is rejected by the incumbents. 

It is conceivable that three non-member countries attempt to form the second FTA. 
In what follows, we argue that they actually have an incentive to form the second FTA. 

We assume that after the first FTA with five countries form, three countries 6, 7, and 8 play the sequential participation game under the open-access rule with consent. 
We evaluate countries' welfare under the following three FTA regimes (ignoring permutations):
\begin{enumerate}[(i)]
    \item $(\{1,2,3,4,5\},\{6\},\{7\},\{8\})$,
    \item $(\{1,2,3,4,5\},\{6, 7\},\{8\})$,
    \item $(\{1,2,3,4,5\},\{6, 7,8\})$.
\end{enumerate}
We simply denote these three FTA regimes by $(5, 1, 1, 1)$, $(5, 2, 1)$ and $(5, 3)$, respectively.
Table 2 gives participant's welfare $W_{\alpha}$ of the first FTA, participant's welfare $W_{\beta}$ of the second FTA, and a non-participant's welfare $W_{\gamma}$. 
The computation is given in the supplementary material S.10. 

We observe from Table 2 that participant's welfare of the second FTA increases as the FTA grows. Therefore, similarly to Theorem~4.6, we can show that the three countries $6, 7$ and $8$ form the second FTA $\{6,7,8\}$ under the open-access rule with consent. 

Note that participant's welfare of the first FTA decreases as the second FTA grows. 
However, their welfare is $487.624$, which is still higher than the welfare under the global FTA. 
Thus, the participants of the first FTA are not motivated to create the global FTA by merging with the second FTA, and thus the regime  $(5,3)$ is realized.

Finally, we consider a possibility of non-overlapping FTAs. 
We here examine whether or not a pair of a participant of the first FTA and a participant of the second FTA in the example have an incentive (in the short run) to form a bilateral FTA between themselves. 

A numerical computation shows that participant's welfare of the first FTA is $487.958$ and  participant's welfare of the second FTA is $487.091$ after the creation of their bilateral FTA. 
Since the bilateral FTA improves both participants' welfare than under the FTA regime $(5, 3)$, this means that it is not pairwise stable in a network formation game (Jackson and Wolinsky 1996). 
We further discuss the difference between our approach and the network stability approach in the next section.

\begin{table}
    \centering \small
        \begin{filecontents*}{num_values2.csv}
$m$ , $t_{\alpha}$ , $t_{\beta}$ , $t_{\gamma}$ , $W_{\alpha}$ , $W_{\beta}$ , $W_{\gamma}$ 
"{(5,1,1,1)}" , 0.160475  , - , 0.160178 , 487.942 , - , 486.209 
"{(5,2,1)}"  , 0.160475 , 0.161194 , 0.161435 , 487.837 , 486.537 , 486.104 
"{(5,3)}" , 0.160475 , 0.160953 , - , 487.628 , 486.761 , - 
\end{filecontents*}
\csvreader[tabular=c|cccccc,
        head=false,
        table head=\toprule,
        late after line=\\,
        late after first line=\\\midrule,
        table foot=\bottomrule,
        before reading={\catcode`\"=9}
        ]{num_values2.csv}{}{\csvcoli & \csvcolii & \csvcoliii & \csvcoliv & \csvcolv & \csvcolvi & \csvcolvii}
    \caption{The effect of two FTAs at parameters $(n,a,\lambda)=(8,12,36)$. Notations: $m$ (regime), $t_{\alpha}$ (tariff on a participant in the large FTA), $t_{\beta}$ (tariff on a participant in the small FTA), $t_{\gamma}$ (tariff on a non-participant), $W_{\alpha}$ (the large FTA participant's welfare), $W_{\beta}$ (the small FTA participant's welfare), $W_{\gamma}$ (non-participant's welfare).}
    \label{fig:numerical_values2}
\end{table}

\section{Discussion}

\subsection{Open Regionalism}

We discuss our results on the formation of global free trade agreement under open-access rules, comparing with the works by Yi (1996, 2000), Seidmann (2009), and Mukunoki and Tachi (2006).

Yi (1996) studies a custom union (CU) formation in a Cournot oligopoly model with linear demand and costs. 
He presents an open regional game, both with simultaneous moves and with sequential moves.
He shows that the grand CU is a unique (pure strategy) Nash equilibrium outcome under the simultaneous-move open regionalism rule.\footnote{Yi (2000) applies the same model to the FTA case. He shows that the grand FTA may not be a Nash equilibrium due to a free-riding problem.} 

His open regionalism rule is similar to our open-access rule without consent, while our model is a sequential-move game. 
In his simultaneous-move game, countries simultaneously announce  an ``address.'' 
The countries choosing the same address form a CU. 
When all countries announce different addresses, it is interpreted as a situation where every country forms a single-member CU by itself. 

Yi's game enables every country to form a two-member CU with any other country by simply  choosing the same address. 
By contrast, when no country participates in an FTA under our open-access rule, any unilateral deviation by one country does not lead to a two-member FTA. 

The critical difference in equilibrium outcomes between Yi (1996) and us lies in that there are multiple SPE outcomes in his sequential-move open regionalism game, while there is a unique SPE outcome, namely the global FTA, in our game. 
A peculiar property of his address-announcing game causes multiple SPE outcomes.

Consider the following three-country example. 
The welfare of each country under a possible FTA regime is given as follows: 
(i) when there is no FTA, every country obtains payoff zero, (ii) when there is a two-member FTA, each participant obtains payoff 2 and a non-participant obtains payoff $-1$, (iii) when there is the global FTA, all members obtain payoff 1. 
Our open-access rule without consent has a unique SPE where three countries join the global FTA. The open-access rule with consent has a unique SPE where the first two countries participate in an FTA and reject the participation of the third country. 

By contrast, Yi's sequential-move open regionalism game has two SPEs. In one SPE, three countries choose the same address, leading to the grand FTA. Whatever address the first country chooses, two succeeding countries coordinate their choices with the first country's one. As a result, the first country can join the global FTA, no matter what address it chooses. In the other SPE, the last two countries form a two-member FTA. Whatever address the first country chooses, two succeeding countries choose the same address, different from the first country's one. The last two countries form a two-member FTA without the first country, regardless of its choice. A coordination problem arises in Yi's sequential model.

Seindmann (2009) considers another sequential model for trade liberalization under an open-access rule with consent in a three-country setup. In contrast to Yi's (1996) paper and ours, he allows utility transfer and the formation of a bilateral FTA, a bilateral CU, two overlapping FTAs, as well as the global FTA. PTAs can be renegotiated toward the global FTA. 
He demonstrates that patient enough countries may initially form a two-member PTA (possibly a hub-and-spoke regime), which will grow to the global FTA. 
His result reveals a different motive of a PTA from ours caused by utility transfer. PTA members strategically manipulate the status quo, seeking advantageous positions in future negotiations (strategic positioning).

Finally, Mukunoki and Tachi (2006) study a sequential bilateral FTA formation game in a three-country oligopoly model with exogenous tariffs, where all firms produce homogeneous goods. 
Each pair of two countries negotiate for a bilateral FTA, or a link, in a fixed order in their game. 
They show that each country has an incentive to become a hub and that this incentive leads to global FTA as a unique Markov-perfect equilibrium when countries are not patient. 
We study a different concern of how a multilateral FTA participation rule affects an equilibrium trade regime, and show that global FTA and a single smaller FTA uniquely prevail, depending on the need for incumbent members' consent.

\subsection{Comparison with Stability Analysis}

The literature on coalition formation has extensively explored two different approaches. 
One is the process-based approach, and the other is the stability-based approach. 
This paper employs the former like the papers which we reviewed in the last subsection. 
The approach (also called a non-cooperative approach) explicitly formulates a coalition formation process and analyzes a subgame perfect equilibrium (and its refinements). 

The stability-based approach, by contrast, applies cooperative solutions by assuming certain types of coalitional behavior and characterizes stable outcomes against coalitional deviations. A coalitional-proof Nash equilibrium (Bernheim et al.\ 1987) and network stability (Jackson and Wolinsky 1996) are popular stability concepts. 
This subsection compares our results with the stability-based approach's results shown by Missios et al.\ (2016), Furusawa and Konishi (2007), and Goyal and Joshi (2006).

Missios et al.\ (2016) consider a coalitional-proof Nash equilibrium of the three-country trade model in this paper. Roughly speaking, an FTA regime is a \emph{coalitional-proof Nash equilibrium} outcome if it is immune to self-enforcing coalitional deviations. 
A coalitional deviation is self-enforcing if it is immune to self-enforcing deviations by \textit{sub-coalitions}. Thus, a self-enforcing coalitional deviation is defined recursively. 
They study an announcement game where each country simultaneously announces other countries' names with whom it wants to form an FTA. In the game, two countries sign a bilateral FTA if they announce each other's names. 

Missios et al.\ (2016) prove that global FTA is a unique coalitional-proof Nash equilibrium outcome. 
Consider the following three-country ($1$, $2$, and $3$) example. 
There are four possible FTA regimes (ignoring permutations): 
$\mathcal{F}^1=(1\text{-}2, 2\text{-}3, 3\text{-}1)$, $\mathcal{F}^2=(1\text{-}2, 3\text{-}1)$, $\mathcal{F}^3=(1\text{-}2, 3)$, $\mathcal{F}^4=(1, 2, 3)$. Notation $i\text{-}j$ means that countries $i$ and $j$ sign a bilateral FTA. 
$\mathcal{F}^1$ means the global FTA, $\mathcal{F}^2$ a hub-and-spoke regime with country 1 being a hub, $\mathcal{F}^3$ a bilateral FTA, and $\mathcal{F}^4$ no FTA. 
The payoff profiles under FTA regimes are given by $W(\mathcal{F}^1)=(3, 3, 3)$, $W(\mathcal{F}^2)=(5, 2, 2)$, $W(\mathcal{F}^3)=(4, 4, -1)$, $W(\mathcal{F}^4)=(0, 0, 0)$.\footnote{The payoff profile under the hub-and-spoke regime $\mathcal{F}^2$ is based on Missios et al.\ (2016).} 

Then the global FTA is a coalitional-proof Nash equilibrium for the following reason. 
Two countries 1 and 2 have a profitable deviation from $\mathcal{F}^1$ to $\mathcal{F}^3$. Country 1, however, has a profitable unilateral deviation from $\mathcal{F}^3$ to $\mathcal{F}^2$, receiving the highest payoff 5. Note that since country 3 announces the name of country 1 as an FTA partner, country 1 can move from $\mathcal{F}^3$ to $\mathcal{F}^2$ by its unilateral deviation. Thus, the coalitional deviation by countries 1 and 2 from $\mathcal{F}^1$ to $\mathcal{F}^3$ is not self-enforcing. This implies that global FTA $\mathcal{F}^1$ is a coalitional-proof Nash equilibrium. 

Under our open-access rule with consent, by contrast, FTA regime $\mathcal{F}^3$ is a unique subgame perfect equilibrium when an order of moves is (1, 2, 3), although $\mathcal{F}^3$ is not a coalitional-proof Nash equilibrium because country 1's unilateral deviation from $\mathcal{F}^3$ to $\mathcal{F}^2$ is self-enforcing.
A key property of a coalitional-proof Nash equilibrium is that it allows self-enforcing deviations only by \textit{sub-coalitions}. 
In the argument above, after $\mathcal{F}^3$ moves to $\mathcal{F}^2$ by country 1's deviation, no further coalitional deviation is allowed by the definition of a coalitional-proof Nash equilibrium. However, countries 2 and 3 has a profitable deviation from $\mathcal{F}^2$ to $\mathcal{F}^1$, coming back to the initial position. Thus, a cycle prevails. 

In our view, it depends on a context whether or not the restriction to self-enforcing deviations by sub-coalitions is reasonable. 
If transaction costs to form an FTA with an outsider are prohibitively high, the restriction may be justified.  

Furusawa and Konishi (2007) and Goyal and Joshi (2006) employ network stability introduced by Jackson and Wolinsky (1996) in an intra-industry trading model. They consider a network of countries where any pair of two countries signing an FTA has a link. 
A network of FTAs is \emph{pairwise stable} if (i) no single country is better off by severing an existing FTA, and (ii) any unlinked pair of countries are better off by creating a new FTA between them. Furusawa and Konishi (2007) prove that the global FTA (the complete network) is a unique pairwise stable network if countries are symmetric and industrial commodities are not highly substitutable. Goyal and Joshi (2006) show a similar result.

The pairwise stability in a network formation game is weak in the sense that the notion considers the above two deviations (i) and (ii) only. 
In particular, the pairwise stability of the global FTA  requires only that no single country is better off by severing a bilateral FTA in the complete network. 
In the three-country example above, the global FTA $\mathcal{F}^1$ is pairwise stable since country $2$ is worse off in the FTA regime $\mathcal{F}^2$ by severing the bilateral FTA $2\text{-}3$. Neither of the other three FTA regimes is pairwise stable. 

The SPE regime $\mathcal{F}^3$ in our open-access game with consent is not pairwise stable because a pair of countries $1$ and $3$ are better-off by creating a new FTA $1\text{-}3$, leading to $\mathcal{F}^2$. However, as we discussed in the case of a coalitional-proof Nash equilibrium, a pair of countries $2$ and $3$ have the incentive to form their bilateral FTA $2\text{-}3$ in $\mathcal{F}^2$, leading to the global FTA $\mathcal{F}^1$. The notion of pairwise stability does not take into account well forward-looking reasoning of negotiating countries, which is critical in a sequential process of trade liberalization.

\subsection{Custom Unions}

We have focused on trade liberalization through FTAs. 
A custom union (CU) is another framework to attain free trade. 
Unlike an FTA, the members of a CU jointly choose their external tariffs. 
Any country member in a CU cannot enter free trade agreements outside it, without consent of other members. 

This subsection only compares tariff setting by FTAs and CUs. 
Suppose that several CUs are formed. 
Let $\mathcal{C}=(C_1, \cdots, C_m)$ be a CU regime. 
Unlike the case of an FTA regime, every two of CUs, $C_i$ and $C_j$, are disjoint. 
The members of each CU choose collectively the external tariffs to maximize the group welfare. 
The internal tariff in any CU is set to be zero. 

Let $c_i$ be the size of each CU $C_i$. 
For each CU $C_l$ where $l=1, \cdots, m$, 
we denote the group welfare of $C_l$ by $W^{C_l} = \sum_{k\in C_l} W_k$. 
For every $i \in C_l$ and $j \notin C_l$, 

\begin{equation}
\frac{\partial W^{C_{l}}}{\partial t_{ij}}
= -2t_{ij} + \frac{2c_{l}}{2n+\lambda}\{a-2p_j^J \} . 
\label{eq:6_1}
\end{equation}
Thus, the first-order condition for $t_{ij}$ to maximize the group welfare $W^{C_l}$ is given by 
\begin{align}
t_{ij} = \frac{c_{l}}{2n+\lambda}(a-2p_j^J) . 
\label{eq:6_2}
\end{align}
Due to the symmetry, \eqref{eq:6_2} shows that every member $i$'s external tariff $t_{ij}$ on non-member $j$ is identical. 

\vspace{2ex}
\noindent \textbf{Proposition 6.1.} 
Let $\mathcal{C}=(C_1, ..., C_m)$ be a CU regime, and let $c_l$ be the size of each $C_l$. For every $C_l$, $i \in C_l$ and $j \notin C_l$, the external tariff of member $i$ of $C_{l}$ on non-member $j$ is given by
\begin{equation*}
t^{\mathcal{C}}_{ij}= \frac{ac_{l}\lambda}{\lambda^2+4n\lambda + 4(n^2 - \sum_{\{k\mid j \notin C_k\}} c_{k}^2)} . 
\end{equation*}

The proposition shows a stark difference in countries' tariffs between under an FTA regime and a CU regime.\footnote{The proof is given in the supplementary material S.9.} 
Each CU's external tariff is determined by the sizes of all CUs, while an FTA's one on a given country is determined only by the sizes of FTAs which the country joins.

When an FTA forms, members do not raise their external tariffs on non-members. They choose the Nash equilibrium tariff  $t^{\mathrm{NE}} =  \frac{a\lambda}{\lambda^2+4n\lambda+4(n^2-n+1)} $. 
In the CU case, by contrast, the external tariffs are larger than the Nash equilibrium tariff. 
When a single CU with $m$ members forms, they choose the external tariff $ \frac{a\lambda}{\lambda^2+4n\lambda+4(n^2-m^2-n+m+1)}>t^{\mathrm{NE}}$. 
The external tariff of a CU violates the GATT Article XXIV. 

\section{Conclusion}

We consider the formation of FTAs in a sequential game to contribute to the debate of whether or not it can lead to global free trade. 
The primary finding is that a unique subgame perfect equilibrium critically depends on a participation rule employed by open regional agreements. 
Under the open-access rule with consent, where an accession needs incumbent members' consent, a unique equilibrium regime is not the global FTA. 
Incumbent members reject new participants if the FTA size exceeds the optimal level. 
By contrast, global FTA is a unique equilibrium regime under the open-access rule without consent. 
Our result provides a game-theoretical support for the policy recommendation that WTO should require any PTA to be open to all WTO members.

Recently, the role of new types of agreements, open plurilateral agreement (OPA) and critical mass agreement (CPA), are discussed as a novel vehicle for groups of countries to promote shared interests outside trade agreements (Hoekman and Sabel 2019). These new agreements are open to all WTO members. Game theoretical analyses of new frameworks within WTO are promising for future works. 

\newpage

\appendixpage

\setcounter{section}{0}
\renewcommand{\thesection}{A.\arabic{section}}
\setcounter{equation}{0}
\counterwithout{equation}{section}
\renewcommand{\theequation}{A.\arabic{equation}}

\section{Proof of Lemma~2.1}

Substituting \eqref{eq:2_2} and \eqref{eq:2_3} into \eqref{eq:2_6} yields
\begin{equation}
\sum_{j=1}^n(a-p_j^I)=\sum_{j=1}^n \lambda_j^I p_j^I . \label{Aeq:1_1}
\end{equation}
Substituting \eqref{eq:2_4} into \eqref{Aeq:1_1} yields
\begin{equation}
\sum_{j=1}^n(a-p_j^I)=\sum_{j \ne i}p_j^I + (1+ \lambda)p_i^I . \label{Aeq:1_2}
\end{equation}
Substituting \eqref{eq:2_5} into \eqref{Aeq:1_2} yields
\begin{equation}
\sum_{j \ne i}(a-p_i^I-t_{ji}) + a-p_i^I=\sum_{j \ne i}(p_i^I+ t_{ji}) + (1+ \lambda)p_i^I . 
\label{Aeq:1_3}
\end{equation}
\eqref{Aeq:1_3} solves 
\begin{equation}
p_i^I = \frac{na - 2\sum_{j \ne i} t_{ji}}{2n + \lambda} . \label{Aeq:1_4}
\end{equation}
By \eqref{Aeq:1_4} and \eqref{eq:2_5}, we have
\begin{equation}
p_j^I=\frac{na - 2\sum_{k \ne i, j} t_{ki} + (2n-2+\lambda)t_{ji}}{2n + \lambda}.
\label{Aeq:1_5}
\end{equation} 
%
Q.E.D.

\section{Proof of Theorem~3.1}

We first prove \eqref{eq:3_3}. 
From \eqref{eq:2_10} and \eqref{eq:2_14}, it holds that
\begin{align*}
\frac{\partial W_i}{\partial t_{ij}}
&= -(a-p_i^J)\frac{\partial p_i^J}{\partial t_{ij}} + p_i^J\frac{\partial p_i^J}{\partial t_{ij}} 
+ a-2p_i^J - 2t_{ij} \frac{\partial p_i^J}{\partial t_{ij}} \\ 
&= (-a+2p_i^J-2t_{ij})\frac{\partial p_i^J}{\partial t_{ij}} + a-2p_i^J \\
&= (-a+2p_i^J-2t_{ij})(1-\frac{2}{2n+\lambda}) + a-2p_i^J \\
&= -2t_{ij} + \frac{2}{2n+\lambda}(a+2t_{ij}-2p_i^J) \\
&= -2t_{ij} + \frac{2}{2n+\lambda}(a-2p_j^J).
\end{align*}
This proves \eqref{eq:3_3}.
From \eqref{eq:2_14} and \eqref{eq:3_4}, we obtain \eqref{eq:3_5}. 
From \eqref{eq:3_5}, $t_{ij}=t_{kj}$ for every pair $i$ and $k$ with $i\neq k$. 
With help of this fact, \eqref{eq:3_5} solves \eqref{eq:3_6}. 
Q.E.D. 

\section{Proof of Theorem~3.2} 

\noindent\textbf{Part (i)}: 
The system of the first order conditions \eqref{eq:3_10} to maximize the total welfare of the world has a unique solution $t_{ij}=0$ for every $i,j$ ($i\neq j$). 
Thus, global FTA with zero tariffs is the unique optimal regime.

\noindent\textbf{Part (ii)}: 
For every $i,j$ ($i \neq j$), since $t_{ij}=0$, it follows from \eqref{eq:2_10} that 
\begin{align}
p_i^I= p_i^J= \frac{na}{2n+\lambda} . 
\label{Aeq:3_1}
\end{align}

Substituting \eqref{Aeq:3_1} into \eqref{eq:2_14} yields 
\begin{align*}
W_i(t) 
&= \frac{n}{2} \left[ (a-\frac{na}{2n+\lambda})^2 + (\frac{na}{2n+\lambda})^2 \right] + \frac{\lambda}{2} (\frac{na}{2n+\lambda})^2 \\ 
&= \frac{na^2}{2} - na \frac{na}{2n+\lambda} + \frac{2n+\lambda}{2}\frac{n^2a^2}{(2n+\lambda)^2}  \\
&= \frac{na^2(2n+\lambda)^2+n^2a^2(2n+\lambda)-2n^2a^2(2n+\lambda)}{2(2n+\lambda)^2} \\ 
&= \frac{na^2(n+\lambda)}{2(2n+\lambda)} . 
\footnotemark
\end{align*}
\footnotetext{The welfare of every country under global FTA can be also derived from the general formula \eqref{eq:4_4} of the welfare of a participant in an FTA in Lemma~4.2 (by setting $m=n$).}

\noindent\textbf{Part (iii)}: 
We prove this part by Lemma~4.2, which gives a general formula \eqref{eq:4_5} of the welfare of a non-participant in an FTA. 
The proof of the lemma is given in the supplementary material S.5. 

Let $t$ be a common tariff for all countries. 
Letting $m=1$ in \eqref{eq:4_5} of Lemma~4.2, we obtain 
%
\begin{align}
& rW^1_{np}(t) - na^2(n+\lambda)(2n+\lambda) \notag \\
&= -2[\lambda^2+4n\lambda+8n-4+(n-2)\lambda^2-2(2n-n^2+1)\lambda+4(n^2-3n+1)] t^2 \notag \\ 
&= -2(n-1)[\lambda^2+2n\lambda +2\lambda+4n]t^2 \notag \\ 
&= -2(n-1)(2+\lambda)(2n+\lambda)t^2 ,  \label{Aeq:3_2} 
\end{align}
where $r=2(2n+\lambda)^2$. 
Substituting $t=t^{\mathrm{NE}}$ into \eqref{Aeq:3_2} proves the part (iii).  
Q.E.D.

\section{Proof of Theorem 4.2} 
\textbf{Part i}: 
First suppose regime $\mathcal{F}^0$, where no FTA is formed. 
Then, all countries impose the common Nash tariff $t^{\mathrm{NE}}>0$ in \eqref{eq:3_6} on every country. 
Substituting \eqref{eq:3_6} into \eqref{eq:2_10} yields the domestic price of good $I$ in country $j$, given by
\begin{align}
p_j^{I,\mathcal{F}^0} = p_j^{I,\mathrm{NE}} 
&= \frac{na + (2+\lambda) t^{NE}}{2n + \lambda} \notag \\ 
&= \frac{na}{2n + \lambda} + \frac{a\lambda(2+\lambda)}{(2n+\lambda)[\lambda^2+4n\lambda+4(n^2-n+1)]} . \label{Aeq:4_1}
\end{align}

Next, suppose regime $\mathcal{F}^i$, where $i\in F$ and $F$ has $m$ members. 
Take $j\in F$ and $k \notin F$. 
From Theorem~4.1, country $k$ imposes the external tariff $t_{ki}^{\mathcal{F}^i}$ on country $i$, given by
\begin{equation*}
t_{ki}^{\mathcal{F}^i} = \frac{a\lambda}{\lambda^2 + 4n\lambda + 4(n^2-(n-m))} >0. 
\end{equation*}
Since $t_{ji}^{\mathcal{F}^i}=0$ for all $j\in F$, substituting it into \eqref{eq:2_10} yields the domestic price of good $I$ in country $j$ imported from country $i$, given by
\begin{align}
p_j^{I,\mathcal{F}^i} 
&= \frac{na - 2(n-m) t_{ki}^{\mathcal{F}^i}}{2n + \lambda} \notag \\
&= \frac{na}{2n + \lambda} - \frac{ 2a\lambda(n-m)}{(2n+\lambda)[\lambda^2 + 4n\lambda + 4(n^2-n+m)]} . \label{Aeq:4_2}
\end{align}

By $n>m$, from \eqref{Aeq:4_1} and \eqref{Aeq:4_2}, 
\begin{equation*}
p_j^{I,\mathcal{F}^i} - p_j^{I,\mathcal{F}^0} 
= -\frac{1}{2n+\lambda}[ 2(n-m)t_{ki}^{\mathcal{F}^i} + (2+\lambda)t^{\mathrm{NE}}]<0.
\end{equation*}

From \eqref{eq:2_2}, \eqref{eq:2_3}, and \eqref{eq:2_7}, import $m_j^I$ of good $I$ in country $j (\neq i)$ is given by $m_j^{I,\mathcal{F}} = a - 2p_j^{I,\mathcal{F}}$ for every regime $\mathcal{F}$.  
Thus, 
\begin{equation}
m_j^{I,\mathcal{F}^i} - m_j^{I,\mathcal{F}^0} = 2 (p_j^{I,\mathcal{F}^0} - p_j^{I,\mathcal{F}^i}) >0 . \label{Aeq:4_3}
\end{equation}

\noindent\textbf{Part ii}: 
Similarly, from \eqref{eq:2_10}, the domestic price of good $I$ in country $k$ is given by
\begin{align}
p_k^{I,\mathcal{F}^i} 
&= \frac{na +(2m+\lambda) t_{ki}^{\mathcal{F}^i}}{2n + \lambda} \notag \\ 
&= \frac{na}{2n + \lambda} + \frac{a\lambda(2m+\lambda)} {(2n+\lambda)[\lambda^2 + 4n\lambda + 4(n^2-n+m)]}  . \label{Aeq:4_4}
\end{align}
From \eqref{Aeq:4_1} and \eqref{Aeq:4_4}, 
\begin{align*}
p_k^{I,\mathcal{F}^i}  - p_k^{I,\mathcal{F}^0} 
&=\frac{a\lambda}{2n + \lambda}\left[ \frac{2m+\lambda}{\lambda^2 + 4n\lambda + 4(n^2-n+m)} - \frac{2+\lambda}{\lambda^2 + 4n\lambda + 4(n^2-n+1)} \right] \\ 
&= \frac{a\lambda}{2n + \lambda} \left[ \frac{(2m-2)(\lambda^2+4n\lambda+4n^2-4n)+(2m+\lambda)-m(2+\lambda)}{[\lambda^2 + 4n\lambda + 4(n^2-n+m)][\lambda^2 + 4n\lambda + 4(n^2-n+1)]} \right] \\ 
&= \frac{a\lambda}{2n + \lambda} \left[ \frac{(m-1)[2\lambda^2+(8n-1)\lambda+8n^2-8n]}{[\lambda^2 + 4n\lambda + 4(n^2-n+m)][\lambda^2 + 4n\lambda + 4(n^2-n+1)]} \right] \\
&>0 .
\end{align*}
Since $p_k^{I,\mathcal{F}^i} > p_k^{I,\mathcal{F}^0}$, the import $m_k^{I,\mathcal{F}^i} < m_k^{I,\mathcal{F}^0}$ from \eqref{Aeq:4_3}. 
Q.E.D. 

\section{Proof of Theorem 4.3} 
By \eqref{eq:4_4}, 
\begin{align}
rW_p^{m+1}(t) 
=& na^2 (n+\lambda)(2n+\lambda) + 4(a\lambda)^2(2m+2+\lambda)(n-m-1)^2 (t'_p)^2 \notag \\
& - 4(a\lambda)^2(n-m-1)^2 t'_p + 4(a\lambda)^2 (n-m-1)(n-1) t'_{np} \notag \\
& - 2(a\lambda)^2(n-m-1) \left[\lambda^2+4n\lambda +8n-4 \right](t'_{np})^2 , 
\label{Aeq:5_1}
\intertext{where }
t'_p &= \frac{1}{\lambda^2+4n\lambda+4(n^2-n+m+1)}  \label{Aeq:5_2} \\ 
t'_{np} &= \frac{1}{\lambda^2+4n\lambda+4(n^2-n+1)} .  \label{Aeq:5_3} 
\end{align}

By \eqref{eq:4_5}, 
\begin{align}
rW^m_{np}(s) &= na^2(n+\lambda)(2n+\lambda) -2m(a\lambda)^2\left[\lambda^2+4n\lambda+8mn-4m^2\right](s'_p)^2 \notag \\
& -2(a\lambda)^2\left[(n-m-1)\lambda^2-2(2mn-n^2+1)\lambda+4(n^2-2mn-n+m)\right](s'_{np})^2 \notag \\
& +4(a\lambda)^2 m(n-m) s'_p - 4(a\lambda)^2 m(n-1) s'_{np} , 
\label{Aeq:5_4}
\intertext{where } 
s'_{p} &= \frac{1}{\lambda^2+4n\lambda+4(n^2-n+m)} \label{Aeq:5_5} \\ 
s'_{np} &= \frac{1}{\lambda^2+4n\lambda+4(n^2-n+1)} .  \label{Aeq:5_6}  
\end{align}
Note that $t'_p<s'_p<t'_{np}=s'_{np}$.

From \eqref{Aeq:5_1} and \eqref{Aeq:5_4}, 
\begin{align}
&\frac{r [W^{m+1}_p (t) - W^m_{np}(s) ]}{(a\lambda)^2} \notag\\
&= 4(2m+2+\lambda)(n-m-1)^2 (t'_p)^2 
- 4(n-m-1)^2t'_p \notag \\
& - 2(n-m-1)\left[\lambda^2+4n\lambda +8n-4 \right](t'_{np})^2 \notag \\
& + 4(n-m-1)(n-1)t'_{np} \notag \\ 
& +2m\left[\lambda^2+4n\lambda+8mn-4m^2\right](s'_p)^2 
- 4 m(n-m)s'_p \notag \\
& +2\left[(n-m-1)\lambda^2-2(2mn-n^2+1)\lambda+4(n^2-2mn-n+m)\right](s'_{np})^2 \notag \\
& +4m(n-1)s'_{np} ,  \label{Aeq:5_7}
\end{align}
where $t'_p, t'_{np}, s'_{p}, s'_{np}$ are given by \eqref{Aeq:5_2}--\eqref{Aeq:5_3} and \eqref{Aeq:5_5}--\eqref{Aeq:5_6}.

We can show that the RHS of \eqref{Aeq:5_7} is strictly positive. 
Since its derivation is tedious, we provide it in the supplementary material S.6. 
Hence $W^{m+1}_p (t) > W^m_{np}(s)$.
Q.E.D.

\section{Proof of Theorem 4.4} 

By Lemma 4.2, 
\begin{align}
\frac{rW_{np}^m(t)}{(a\lambda)^2}
&= \frac{n(n+\lambda)(2n+\lambda)}{\lambda^2} \notag \\
&-2m\left[\lambda^2+4n\lambda+8mn-4m^2\right](t'_p)^2 \notag \\
&-2\left[(n-m-1)\lambda^2-2(2mn-n^2+1)\lambda+4(n^2-2mn-n+m)\right](t'_{np})^2 \notag \\
&+4m(n-m)t'_p-4m(n-1)t'_{np} , \label{Aeq:6_1}
\end{align}
where $t'_p$ and $t'_{np}$ are given in \eqref{Aeq:5_2} and \eqref{Aeq:5_3}. 

Let function $f$ be the RHS of \eqref{Aeq:6_1} except the first constant term. 
By $\frac{\partial t_{np}'}{\partial m}=0$, 
\begin{align}
\frac{\partial f(m)}{\partial m}
=& -2\left[\lambda^2+4n\lambda+16mn-12m^2\right](t'_p)^2 \notag \\
& -4m\left[\lambda^2+4n\lambda+8mn-4m^2\right]t'_p \frac{\partial t'_p}{\partial m} \notag\\
& +2\left[\lambda^2+4n\lambda+8n-4\right](t'_{np})^2 \notag \\
& +4(n-2m)t'_p + 4m(n-m)\frac{\partial t'_p}{\partial m} -4(n-1)t_{np}. \label{Aeq:6_2}
\end{align}

Furthermore, substituting $\frac{\partial t'_p}{\partial m}=-\frac{4}{(t'_p)^2}$ into \eqref{Aeq:6_2} yields  
\begin{align}
\frac{\partial f(m)}{\partial m}= 
& -2\left[\lambda^2+4n\lambda+16mn-12m^2\right](t'_p)^2 \notag \\
& +16m\left[\lambda^2+4n\lambda+8mn-4m^2\right]\frac{1}{t'_p} \notag\\
& +4(n-2m)t'_p -16m(n-m)\frac{1}{(t'_p)^2} \notag\\
& +2\left[\lambda^2+4n\lambda+8n-4\right]t_{np}^2-4(n-1)t_{np} . \label{Aeq:6_3}
\end{align}

We can show that the derivative $\frac{\partial f(m)}{\partial m}$ in $m$ is strictly negative.  
Since its computation is simple but tedious, we provide it in the supplementary material S.7. 
Hence non-participant country's welfare is decreasing in $m$ with $1\le m \le n-1$.
Q.E.D.

\section{Proof of Theorem~4.5}

Without any crucial loss of generality, we assume that $W_p^m$ has a unique maximum point $m^*$.\footnote{If there are multiple maximum points, then we choose the largest one.} 
Recall that $n$ countries sequentially decide to participate in an FTA or not, according to the order $(1, 2, \cdots, n)$.

We prove the theorem by backward induction in the following three steps.

\noindent \textit{Step 1.} 
First, consider the $n$-th stage game when there are any $m$ incumbents of an FTA with $m<n$.\footnote{If $m^*=n$, then this case is vacuous.} 
If country $n$ participates, then all $m$ participants either accept or reject $n$'s participation independently. 
If the participation is accepted, then every participant $i$ receives payoff $W_p^{m+1}$. 
Otherwise, it receives payoff $W_p^{m}$. 

When $W_p^{m+1}\geq W_p^{m}$, it is (weakly) dominant for every incumbent to accept $n$'s participation.\footnote{If an incumbent is indifferent between accepting and rejecting a new participant, we assume the tie-breaking rule to accept it.} 
Country $n$ optimally participates in an FTA. 
When $W_p^{m+1}< W_p^{m}$, it is (weakly) dominant for every incumbent to reject $n$'s participation. Country $n$'s choice does not affect the game's outcome. 

By the same argument, we can characterize the equilibrium actions for incumbents and a new participant in every $t$-th stage game where $t=m^*+1, \cdots, n$ with $m$ incumbents as follows.
All incumbents accept the participation of country $t$ if and only if $v(m, t) \geq W_p^m$, and country $t$ participates if $v(m, t) \geq W_p^m$, where $v(m, t)=\max \{W_p^{m+1}, \cdots, W_p^{m+n-t+1}\}$. 
\vspace{2ex}

\noindent \textit{Step 2.} 
Next, consider the $m^*$-th stage game when there are $m^*-1$ incumbents of an FTA. 
If all incumbents accept the participation of $m^*$, then they receive payoff $W_p^{m^*}$. 
By $W_p^{m^*}>W_p^{m}$ for all $m\ne m^*$, it is (weakly) dominant for every incumbent to reject country $m^*$'s participation. 
Then, it is also optimal for country $m^*$ to participate in the FTA.
\vspace{2ex}

\noindent \textit{Step 3.}
Finally, consider the $(m^*-1)$-th stage game when there are $m^*-2$ incumbent members of an FTA. 
If all incumbents accept the participation of $m^*-1$, then they receive payoff $W_p^{m^*}$ because all countries rationally expect the size of an FTA expands to $m^*$ in the next $m^*$-th stage.  
By $W_p^{m^*}>W_p^{m}$ for all $m\ne m^*$, it is (weakly) dominant for every incumbent to accept country $m^*-1$'s participation. 
Then, country $m^*-1$ optimally participates in the FTA. 
\vspace{2ex}

By backward induction, the above steps show that all countries $i=1, \cdots, m^*$ optimally participates in an FTA, and that their participations are accepted. 
After country $m^*$ participates, no more countries are accepted. 
Q.E.D.

\section{Proof of Theorem~4.6}

We first show that the equilibrium FTA size $m^*$ is strictly less than $n$.
From \eqref{eq:4_4},
\begin{align*}
  rW_{p}^n =& na^2(n+\lambda)(2n+\lambda) \\ 
  rW_{p}^{n-1}(t) =& na^2(n+\lambda)(2n+\lambda) + 4(2(n-1)+\lambda)(n-(n-1))^2 t_{p}^2 \notag \\
    & - 4a\lambda(n-(n-1))^2t_{p} + 2(n-(n-1))\left[4-8n-4n\lambda - \lambda^2 \right]t_{np}^2 \notag \\
    & + 4a\lambda (n-(n-1))(n-1)t_{np}. 
\end{align*}

From \eqref{Aeq:5_2}--\eqref{Aeq:5_3}, $t_p=a\lambda t_p'$, $t_{np}=a\lambda t_{np}'$, and the difference  
\begin{align}
  r[W_p^{n-1}(t) - W_{p}^n] &= 4(2(n-1)+\lambda) t_{p}^2 - 4a\lambda t_{p} \notag \\
  & \:\:\:\:\:  - 2\left[-4+8n+4n\lambda + \lambda^2 \right]t_{np}^2 + 4a\lambda (n-1)t_{np} \notag \\
  &= 4(a\lambda t'_p)^2 [2(n-1)+\lambda - (2n+\lambda)^2 +4] \notag \\
  & \:\:\:\:\:  + 2(a\lambda t'_{np})^2 [2(n-1)(2n+\lambda)^2-8(n-1)^2+4-8n-4n\lambda - \lambda^2] \notag \\
  &= 2(a\lambda t'_p)^2 [4(n+1)+2\lambda - 2(2n+\lambda)^2 ] \notag \\
  & \:\:\:\:\:  + 2(a\lambda t'_{np})^2 [2(n-1)(2n+\lambda)^2-8(n-1)^2 +4(n-1)^2 -(2n+\lambda)^2] \notag \\ 
  &= 2(a\lambda t'_p)^2 [4(n+1)+2\lambda - 2(2n+\lambda)^2 ] \notag \\  
  & \:\:\:\:\:  + 2(a\lambda t'_{np})^2 [(2n-3)(2n+\lambda)^2-4(n-1)^2] . \label{Aeq:9_1} 
\end{align}

Let function $\zeta(n) = [(2n-3)(2n+\lambda)^2-4(n-1)^2]$. 
For $n\geq 3$, $\zeta(n)>0$ because $\zeta(3)=3(6+\lambda)^2-16>0$ and $d \zeta(n) /d n = 2\lambda^2+ 4\lambda(4n-3) +8n(3n-4) +8>0$. 

Thus, for $n\ge 3$, by $t'_p<t'_{np}$ and \eqref{Aeq:9_1}, 
\begin{align*}
  r[W_p^{n-1}(t) - W_{p}^n] 
  &> 2(a\lambda t'_p)^2 [4(n+1)-4(n-1)^2+(2n+\lambda)^2(2n-5)] \\ 
  &= 2(a\lambda t'_p)^2 [-4n(n-3)+4n^2(2n-5)+\lambda(4n+\lambda)(2n-5)] \\ 
  &= 2(a\lambda t'_p)^2 [4n(2n^2 -6n +3) +\lambda(4n+\lambda)(2n-5)] \\ 
  &> 0 .
\end{align*}

The last inequality holds true because both $(2n^2 -6n +3)$ and $(2n-5)$ are strictly positive. 
Thus, $W_p^{n-1} > W_{p}^n(t)$ for $n\ge 3$, and $W_p^m$ is not maximized at $n$, i.e.\ $m^*<n$. 

Next, we show that when $n$ is sufficiently large, 
welfare $W_p^m$ is maximized at $m^*=(n/2)-1$, $n/2$, or $(n/2)+1$ for even $n$, 
and at $m^*=(n-1)/2$ or $(n+1)/2$ for odd $n$. 
Let $\alpha = m/n\in (0,1]$ and fix $n$. 
By \eqref{Aeq:5_2}--\eqref{Aeq:5_3}, $\frac{\partial t'_p}{\partial \alpha}=-4nt_p^2$ and $\frac{\partial t'_{np}}{\partial \alpha}=0$. 
Then, by \eqref{eq:4_4}, the derivative of $W_p^m$ with respect to $\alpha$ is given by 
\begin{align*}
\frac{\partial W_p^m}{\partial \alpha} 
&= 8(a\lambda)^2 n^2(1-\alpha)t'_p - 4(a\lambda)^2 n^2 t'_{np} + R(n) \\
&\to (a\lambda)^2(1-2\alpha) \text{ as $n \to \infty$}, 
\end{align*}
where $R(n)$ is the remainder term.
$R(n)$ satisfies $\lim_{n\to \infty} R(n) =0$ because $t'_p,t'_{np}$ have the order of $1/n^2$.
Since it is tedious, the derivation is given in the supplementary material S.8. 

Since the derivative is monotonically decreasing in $\alpha=\frac{m}{n}\in (0,1]$, welfare function $W_p^m$ is concave in $m$ in the limit. 
When $\alpha=1/2$, $\lim_{n\to \infty} \frac{\partial W_p^m}{\partial \alpha}= 0$. 
Therefore, for sufficiently large $n<\infty$, welfare $W_p^m$ is maximized at $m^*=(n/2)-1$, $n/2$, or $(n/2)+1$ for even $n$ and at $m^*=(n-1)/2$ or $(n+1)/2$ for odd $n$. 
The remainder term $R(n)$ determines which maximizes the welfare. 
Q.E.D.

\section{Proof of Theorem~4.7} 
We prove the theorem by backward induction. 
First, consider the optimal choice of the last country. 
From Theorem~4.3, it is optimal for the last country to participate in an FTA if there exists at least one incumbent, who has participated in the FTA before. 
Otherwise, it is indifferent between participation and non-participation. 
If they are indifferent, we assume the tie-breaking rule under which the last country participates.

Next, consider the optimal choice of the second-to-last country. 
Again, from Theorem~4.3, it is optimal for the second-to-last country to participate in an FTA if there exists at least one incumbent, who has participated in the FTA before. 

Suppose that there have been no participation in an FTA. 
If country $i$ participates, then it receives the welfare $W_p^2$, given the optimal choice of the last country. 
If country $i$ does not participate, then it receives the welfare $W_{np}^1$, regardless of the indifferent choices of the last country. 

Since $W_p^2>W_{np}^1$ by Theorem~4.3, it is also optimal for country $i$ to participate in this case. 
Since the second-to-last country always participates, by Theorem~4.4, all other countries optimally participate in an FTA at their all moves.

Thus, global free trade is a unique SPE. 
The argument holds independently of an order of moves. 
Q.E.D.


\end{document}